\documentclass{ieeeaccess}
\usepackage{cite}
\usepackage{amsmath,amssymb,amsfonts}
\usepackage{graphicx}
\usepackage{textcomp}
\usepackage{dcolumn}
\usepackage{bm}
\usepackage{mathtools}
\usepackage{braket}
\usepackage{enumitem}
\usepackage{empheq}
\usepackage{cancel}
\usepackage[caption=false,font=footnotesize]{subfig}
\usepackage{hyperref}
\usepackage{cuted}
\usepackage{capt-of}
\usepackage{lettrine}
\usepackage[normalem]{ulem}
\usepackage{amsthm}
\newtheorem{theorem}{Theorem}
\newtheorem{corollary}{Corollary}
\usepackage{algorithm}
\usepackage{algpseudocode}

\newcommand{\ir}{\Omega_{\mathrm{min}}}
\newcommand{\uv}{\Omega_{\mathrm{max}}}

\providecommand{\xfigwd}{0pt}

\def\BibTeX{{\rm B\kern-.05em{\sc i\kern-.025em b}\kern-.08em
    T\kern-.1667em\lower.7ex\hbox{E}\kern-.125emX}}

\renewcommand{\EOD}{\global\eoddefinedtrue}

\begin{document}

\history{Date of publication xxxx 00, 0000, date of current version xxxx 00, 0000.}
\doi{10.1109/TQE.2026.DOI}

\title{Bandwidth-Cost Reduction for Photon-Matter Simulation at the Single-Photon Level}

\author{\uppercase{Romain Piron}\authorrefmark{1} and
\uppercase{Akihito Soeda}\authorrefmark{1,2,3}}
\address[1]{National Institute of Informatics, Chiyoda-ku, Tokyo, Japan}
\address[2]{Informatics Program, Graduate Institute for Advanced Studies, SOKENDAI, Tokyo, Japan}
\address[3]{Department of Physics, Graduate School of Science, The University of Tokyo, Japan}
\tfootnote{This work was supported by JST Moonshot R\&D Program Grant Numbers JPMJMS226C and
JPMJMS256K. }



\begin{abstract}
Simulations of quantum dynamics of traveling photons interacting with matter can be used to model and design photonic quantum-network components. Such simulations require selecting a finite set of photonic modes, where smaller sets are more desirable for computational efficiency. In this work, however, we show that opting for smaller subsets alters simulation results, if done naively. We therefore provide a consistency condition that can be used to calibrate the simulation input parameters to increase efficiency while preserving accuracy, simultaneously. This condition is then used to derive a simple prescription that reduces the propagation cost by two order of magnitude in a waveguide benchmark. Finally, we extend the same calibration principle to a two-dimensional atomic array, demonstrating the simultaneous benefits beyond analytically solvable waveguide models.
\end{abstract}


\maketitle

\section{Introduction}
\label{sec:introduction}

\lettrine{P}hoton-matter interactions play a crucial role in quantum information technologies, with applications ranging from photonic quantum computing~\cite{knillSchemeEfficientQuantum2001,kokReviewArticleLinear2007a} to quantum networking~\cite{kozlowskiArchitecturalPrinciplesQuantum2023a, wehnerQuantumInternetVision2018}. As these technologies fundamentally rely on the transmission of photonic quantum states, the susceptibility of these states to imperfections arising during propagation and interactions with matter represents a critical challenge for quantum technologies. 

In the context of quantum networks, the seminal quantum teleportation protocol~\cite{bennettTeleportingUnknownQuantum1993}, whose principles were subsequently extended to enable nonlocal quantum operations between spatially separated processors in distributed quantum computation~\cite{ciracDistributedQuantumComputation1999, eisertOptimalLocalImplementation2000}, established shared entanglement as a fundamental resource. To generate such entanglement between distant nodes, photons serve as natural carriers of quantum information, acting as flying qubits in remote entanglement-generation protocols~\cite{barrettEfficientHighfidelityQuantum2005,cabrilloCreationEntangledStates1999,duanEfficientEngineeringMultiatom2003}. Hence, imperfections in the photonic states can ultimately compromise the quality of networked quantum operations. 

This is particularly relevant to distributed quantum computation~\cite{vanmeterPathScalableDistributed2016}, where distortions of the photonic states used for remote entanglement generation can reduce the fidelity of the resulting Bell pairs and, in turn, introduce errors into quantum operations performed across distant processors~\cite{rametteFaulttolerantConnectionErrorcorrected2024a,sinclairFaulttolerantOpticalInterconnects2025}. Understanding how these network-level errors originate from imperfections in photonic states therefore calls for a quantum-mechanical description that explicitly tracks their evolution and interactions with network components, including quantum memories, repeaters, and detectors. This perspective calls for theoretical frameworks in which localized matter degrees of freedom are coupled to propagating quantized electromagnetic fields~\cite{ramosNonMarkovianDynamicsChiral2016,combesSLHFrameworkModeling2017}. By retaining the underlying quantum dynamics, such approaches provide a means to investigate the microscopic mechanisms responsible for photonic-state imperfections and their consequences for network performance. 

To describe the underlying photon-matter interactions at the microscopic level, a Hamiltonian provides a natural description of the quantum evolution of propagating photons interacting with localized matter systems. In such models, the relevant matter degrees of freedom are often represented as effective two-level systems (TLSs)~\cite{scullyQuantumOptics1997,cohen-tannoudjiPhotonsAtomsIntroduction1989}. When photon propagation can be approximated as one-dimensional, as is often the case when modeling light propagation in optical fibers, localized TLSs couple to photons propagating along a single spatial coordinate. This setting, commonly referred to as waveguide quantum electrodynamics (QED), has been extensively studied theoretically~\cite{sheremetWaveguideQuantumElectrodynamics2022,royStronglyInteractingPhotons2017,rouletTwoPhotonsAtomic2016,greenbergSinglephotonScatteringQubit2023,domokosQuantumDescriptionLight2002,shenCoherentSinglePhoton2005,shenStronglycorrelatedMultiparticleTransport2007,shenStronglyCorrelatedTwoPhoton2007,fanInputOutputFormalismFewPhoton2010,pletyukhovScatteringMasslessParticles2012,chenPropagatorsJaynesCummingsModel2002,ralleyHongOuMandelInterferenceSingle2015,shiMultiphotonScatteringTheory2015,schneiderGreensfunctionFormalismWaveguide2016, nysteenScatteringTwoPhotons2015}, with theoretical predictions also supported by experimental observations~\cite{lauchtWaveguideCoupledOnChipSinglePhoton2012,astafievResonanceFluorescenceSingle2010}. Beyond waveguide-QED, quantum photon-matter models are not limited to effectively one-dimensional propagation. Higher-dimensional geometries have also been considered~\cite{gonzalez-tudelaMarkovianNonMarkovianDynamics2017,perczelTopologicalQuantumOptics2017,martinezMixedPerturbativenonperturbativeTreatment2024}. In particular, two-dimensional models based on ensembles of TLSs have been used to construct microscopic descriptions of optical elements~\cite{havukainenQuantumSimulationsOptical1999,obaFastSimulationMultiphoton2024}. In the latter direction, such models have been used to describe mirrors at the single-photon level~\cite{havukainenQuantumSimulationsOptical1999}, and have more recently been extended to multi-photon simulations of quantum-optical phenomena~\cite{obaFastSimulationMultiphoton2024} such as Hong-Ou-Mandel interference~\cite{hongMeasurementSubpicosecondTime1987a} and Bell-CHSH violation~\cite{bellEinsteinPodolskyRosen1964,clauserProposedExperimentTest1969, aspectExperimentalRealizationEinsteinPodolskyRosenBohm1982}. However, moving beyond simplified one-dimensional settings or considering more complex photon-matter systems rapidly makes analytical solutions difficult to obtain, therefore making numerical approaches necessary to study microscopic optical elements and multi-photon interference phenomena~\cite{havukainenQuantumSimulationsOptical1999,obaFastSimulationMultiphoton2024}. This limitation is further exacerbated when the photon-matter coupling has a nontrivial spatial or spectral structure. Precisely tracking the quantum state of propagating photons in such complex scenarios therefore requires numerical simulation methods.

In waveguide-QED systems with effectively frequency-independent photon-matter coupling, corresponding to a Markovian regime, the dynamics can often be described through time-local input-output relations~\cite{fanInputOutputFormalismFewPhoton2010,kiilerichInputOutputTheoryQuantum2019a,combesSLHFrameworkModeling2017}. This structure has enabled efficient time-domain simulation methods based on time-bin discretizations~\cite{heuckPhotonPhotonInteractionsDynamically2020,bundgaard-nielsenWaveguideQEDjlEfficientFramework2025,regidorQwaveMPSEfficientOpensource2026}. These approaches are particularly powerful in one-dimensional Markovian settings, and can also incorporate certain non-Markovian effects by introducing explicit feedback loops~\cite{arranzregidorModelingQuantumLightmatter2021,crowderQuantumTrajectoryTheory2020}. However, the compactness of the time-domain picture becomes less direct when the photon-matter interaction has a nontrivial spectral structure. Moreover, representing propagating photons solely through a time-bin index naturally relies on an effectively one-dimensional propagation geometry. This motivates investigating complementary simulation strategies that keep the photonic modes explicitly in momentum space, in the spirit of microscopic simulations of optical elements built from ensembles of TLSs~\cite{havukainenQuantumSimulationsOptical1999,obaFastSimulationMultiphoton2024}, with the long-term goal of treating broader spectral structures and more complex photon propagation geometries.

Momentum-space photon-matter Hamiltonians are central to many theoretical treatments of quantum photon-matter models, in particular those based on scattering theory and Green-function methods~\cite{chenPropagatorsJaynesCummingsModel2002,schneiderGreensfunctionFormalismWaveguide2016,greenbergSinglephotonScatteringQubit2023,sheremetWaveguideQuantumElectrodynamics2022,martinezMixedPerturbativenonperturbativeTreatment2024}. However, their direct numerical simulation is challenging because the photonic field is described by a continuum of modes, which must be discretized and truncated for numerical treatment. Reducing the size of the resulting set of photonic modes directly decreases the dimension of the numerical problem, thereby reducing the computational cost~\cite{woodsSimulatingBosonicBaths2015,obaFastSimulationMultiphoton2024}. At the same time, such truncations must be handled carefully: observables depend on how the continuum is regularized, and coupling a discrete system to a truncated bosonic bath can induce effective parameter shifts, analogous to renormalization effects in field theory~\cite{peskinIntroductionQuantumField1995,collinsRenormalizationIntroductionRenormalization1984}. Related cutoff-dependent shifts, including Lamb-shift-like corrections, have been discussed in photon-matter interaction models~\cite{malekakhlaghCutofffreeCircuitQuantum2017,sheremetWaveguideQuantumElectrodynamics2022,martinezMixedPerturbativenonperturbativeTreatment2024,greenbergSinglephotonScatteringQubit2023}. This creates a practical tradeoff between computational efficiency and physical accuracy, and calls for a systematic analysis of how numerical simulation of photon-matter interactions depends on the chosen spectral truncation.

In this work, we investigate numerical simulations of photon-matter interactions within a restricted frequency range, using what we refer to as a finite-bandwidth regularization. In practice, this amounts to restricting the photonic continuum to a finite frequency range bounded by infrared (IR) and ultraviolet (UV) cutoffs. By systematically examining how these cutoffs affect the simulated dynamics, we identify the underlying mechanisms responsible for deviations from the desired physical behavior. This analysis leads to a practical cutoff-consistency condition that provides a basis for calibrating effective device parameters, enabling accurate simulations at reduced computational cost. The paper is organized as follows. Sec.~\ref{sec:hamiltonian_model} introduces the Hamiltonian framework and the two model classes used throughout the work. Sec.~\ref{sec:cutoff_regularization} describes the regularization scheme and the associated scaling of the numerical simulation. Sec.~\ref{sec:bandwidth_diagnosis} uses a single-photon waveguide benchmark to diagnose the effect of reducing the frequency window. Sec.~\ref{sec:green_matrix_response} draws on scattering theory to formulate a single-photon consistency condition for reduced-bandwidth simulations. Sec.~\ref{sec:waveguide_calibration} applies this condition to calibrated waveguide simulations and quantifies the resulting computational gain. Sec.~\ref{sec:two_dimensional_calibration} extends the calibration strategy to a two-dimensional atomic array. Finally, Sec. VIII concludes the paper. 

\subsection*{Conventions}

We adopt the following conventions throughout the paper:
\begin{itemize}
    \item We adopt natural units $\hbar=c=1$, express all quantities in units of energy, and omit explicit units unless otherwise stated.
    \item The subscripts $0$ and $A$ denote numerical and target parameters, respectively. For multiple atoms, the atom index follows this label, e.g. $\Omega_{0,j}$ and $g_{A,j}(\bm{p})$. 
    \item For a function $f$ of the complex variable $z$ and real $\omega$, we write
    \begin{align*}
        f(\omega+i0^+) \equiv \lim_{\epsilon\to 0^+} f(\omega+i\epsilon),
    \end{align*}
    i.e., the boundary value approached from the upper half of the complex plane.
    \item For a matrix $\mathsf{X}$, its scalar entries are written $X_{ij}=[\mathsf{X}]_{ij}$.
\end{itemize}
In addition, the main notations used throughout the paper are summarized in the following table. \vspace{0.4cm} \\
\hspace{-0.1cm}
\begin{tabular}{|ll|}
    \hline
    \textit{Object} \hspace{5cm} & \textit{Notation} \\
    \hline
    Scalar quantity & $x$ \\
    Vector & $\bm{x}$ \\
    Matrix & $\mathsf{X}$ \\
    Diagonal matrix such that $X_{ii} = x_i$ & $\operatorname{Diag}(\bm{x})$ \\
    $\ell_2$ norm & $\left\lVert \cdot \right\rVert_2$ \\
    &  \\
    Atomic frequency (numerical) &  $\Omega_{0}$  \\
    Atomic frequency (target) & $\Omega_{A}$ \\
    Coupling function (numerical) & $g_{0}(\bm{p})$ \\
    Coupling function (target) & $g_{A}(\bm{p})$ \\
    Decay rate (numerical)&  $\gamma_{0}$  \\
    Decay rate (target)& $\gamma_{A}$ \\
    Dimensionless coupling (numerical)&  $d_{0}$  \\
    Dimensionless coupling (target) & $d_{A}$ \\
    &  \\
    Simulation time & $T$ \\
    Time step & $\Delta t$ \\
    Number of time steps & $N_{\mathrm{RK4}}$ \\
    Infrared (IR) cutoff & $\ir$ \\
    Ultraviolet (UV) cutoff & $\uv$ \\
    \hline
\end{tabular}

\section{Hamiltonian modeling of photon-matter systems}
\label{sec:hamiltonian_model}

\begin{figure*}
    \centering
    \includegraphics[width=\linewidth]{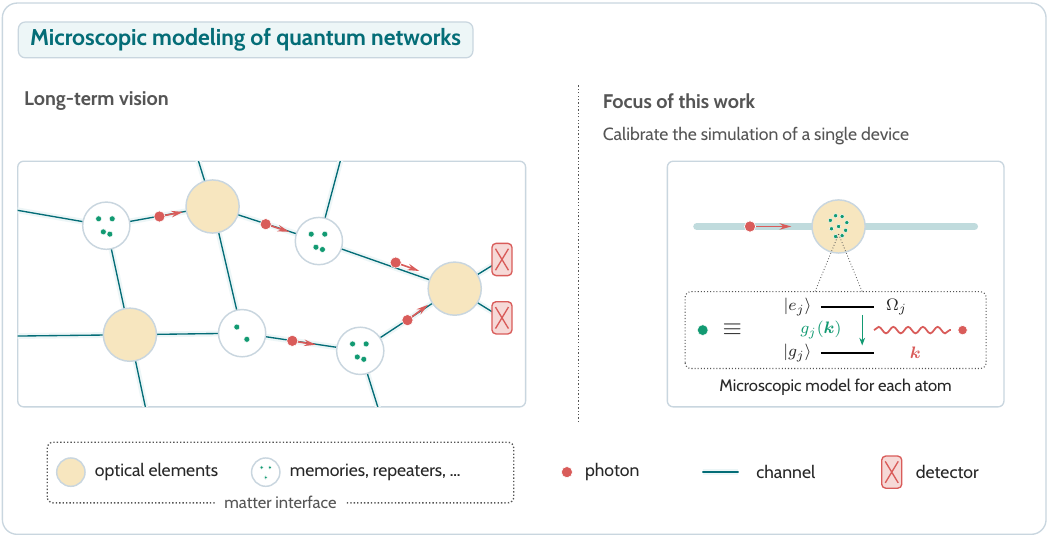}
    \caption{Long-term target and reduced single-device setting studied in this work. Matter interfaces are modeled microscopically as TLS ensembles coupled to photonic modes.}
   \label{fig:qnet_vision}
\end{figure*}

Motivated by the simulation of traveling photons interacting with matter interfaces in quantum networks, as illustrated in Fig.~\ref{fig:qnet_vision}, we consider here a reduced setting consisting of a single elementary device. This provides a controlled starting point for developing and benchmarking the numerical framework. In this section, we introduce the Hamiltonian framework and the two model classes used in the simulations.

\subsection{Continuum photon-matter Hamiltonian}

We consider $N_A$ TLSs coupled to a continuum of photonic modes in $d$ spatial dimensions~\cite{havukainenQuantumSimulationsOptical1999,obaFastSimulationMultiphoton2024,royStronglyInteractingPhotons2017}. The photonic modes are labeled by a momentum $\bm{p}\in\mathbb{R}^d$ and have the dispersion relation $\omega_{\bm{p}}$, which we leave unspecified until introducing the two concrete models considered below. Throughout this work, we use natural units in which $\hbar = c = 1$, so that all dimensionful quantities are expressed in powers of energy. In these units, the free Hamiltonian reads~\cite{havukainenQuantumSimulationsOptical1999,obaFastSimulationMultiphoton2024,royStronglyInteractingPhotons2017}
\begin{align*}
    H_{\mathrm{free}} = \int \frac{d^d\bm{p}}{(2\pi)^d}\, \omega_{\bm{p}}\, a^{\dagger}(\bm{p})a(\bm{p}) + \sum_{j=1}^{N_A} \Omega_{0,j}\,\sigma_j^+\sigma_j^- .
\end{align*}
Here $\Omega_{0,j}$ is the numerical transition frequency of TLS $j$, while
$\sigma_j^+=\ket{e_j}\bra{g_j}$ and $\sigma_j^-=\ket{g_j}\bra{e_j}$ are its raising and lowering operators. The photonic operators satisfy
\begin{align*}
    [a(\bm{p}),a^\dagger(\bm{p}')] = (2\pi)^d\delta^{(d)}(\bm{p}-\bm{p}').
\end{align*}
Within the rotating-wave approximation (RWA), the interaction is of Jaynes-Cummings type~\cite{jaynesComparisonQuantumSemiclassical1963}:
\begin{align*}
    H_I = \sum_{j=1}^{N_A} \int \frac{d^d\bm{p}}{(2\pi)^d} \left[ g_{0,j}(\bm{p})\,a(\bm{p})\,\sigma_j^+ + g_{0,j}(\bm{p})^*\,a^\dagger(\bm{p})\,\sigma_j^- \right].
\end{align*}
The function $g_{0,j}(\bm{p})$ describes the coupling between TLS $j$ and mode $\bm{p}$. A straightforward dimensional analysis shows that, in the continuous formulation, it has dimensions
\begin{align*}
[g_{0,j}(\bm{p})]=[\mathrm{energy}]^{1-d/2}.
\end{align*}
The full Hamiltonian is parameterized with the set of numerical TLS frequencies and couplings:
\begin{align*}
    H\Big[\{\Omega_{0,j}\},\{g_{0,j}\}\Big] = H_{\mathrm{free}}+H_I .
\end{align*}
In particular, the subscript $0$ emphasizes that these are the \emph{numerical} parameters fed into the simulation. The goal of this paper is to discuss how they should be adjusted to reproduce a prescribed response. As we will see, it requires careful analysis when the effective bandwidth of the model is reduced. Finally, for later convenience, we also introduce the spectral-density matrix~\cite{cohen-tannoudjiPhotonsAtomsIntroduction1989, martinezMixedPerturbativenonperturbativeTreatment2024, breuerTheoryOpenQuantum2007}
\begin{align}
    \mathsf{J}(\omega) = \int \frac{d^d\bm{p}}{(2\pi)^d}\, \bm{g}_{0}(\bm{p})\bm{g}_{0}^{\dagger}(\bm{p}) \,\delta(\omega-\omega_{\bm{p}}),
    \label{eq:spectral_density_matrix}
\end{align}
where $\bm{g}_{0}(\bm{p})=(g_{0,1}(\bm{p}),\ldots,g_{0,N_A}(\bm{p}))^{\mathsf{T}}$ is the column vector of couplings. The entries are denoted $J_{jk}(\omega)=[\mathsf{J}(\omega)]_{jk}$. When $N_A = 1$, we simply write $J(\omega)=J_{11}(\omega)$.

\subsection{Fixed-excitation basis}

The total Hilbert space is the tensor product of the photonic Fock space and the atomic Hilbert space,
\begin{align*}
    \mathcal{H} = \mathcal{H}_{\mathrm{photons}} \otimes \mathcal{H}_{\mathrm{atoms}}.
\end{align*}
A natural basis for this problem is provided by the eigenstates of the free Hamiltonian $H_{\mathrm{free}}$. A state with $m$ photons and $q$ atomic excitations, with $J=\{j_1,\dots,j_q\}$ denoting the set of excited atoms, is written compactly as
\begin{align*}
    \begin{split}
        \ket{\bm{p}_1,\dots,\bm{p}_m, e_{j_1}, \dots, e_{j_q}} \phantom{\frac{1}{2}} & \\
        & \hspace*{-11em} = a^\dagger(\bm{p}_1)\cdots a^\dagger(\bm{p}_m) \ket{\emptyset_{\mathrm{photons}}} \otimes \left(\prod_{j \in J} \sigma_j^{+} \right)\ket{g_1, \dots, g_{N_A}}.
    \end{split}
\end{align*}
The following conventions apply:
\begin{itemize}
    \item $q=0$: all atoms are in their ground state, and the state is denoted $\ket{\bm{p}_1,\dots,\bm{p}_m}$.
    \item $m=0$: the photonic field is in the vacuum state, and the state is denoted $\ket{e_{j_1},\dots,e_{j_q}}$.
    \item $m=q=0$: no creation or excitation operators are applied, yielding the global vacuum state. This state will not appear in this paper.
\end{itemize}
Because the rotating-wave Hamiltonian conserves the total number of excitations~\cite{larsonJaynesCummingsModel2021,pletyukhovScatteringMasslessParticles2012},
\begin{align*}
    N_{\mathrm{exc}} = \int \frac{d^d\bm{p}}{(2\pi)^d}\, a^\dagger(\bm{p})a(\bm{p}) + \sum_{j=1}^{N_A}\sigma_j^+\sigma_j^- ,
\end{align*}
the dynamics can be restricted to fixed-excitation sectors. The $N$-excitation sector contains states satisfying $m+q=N$.

\subsection{Model specifications}

As discussed below, the momentum integral must be truncated in numerical simulations. In this paper, we study how to handle this truncation and illustrate our findings using the two models summarized in Fig.~\ref{fig:model_calibration_summary}.

The first model is a one-dimensional waveguide setup, in which a single TLS is positioned at the origin of a one-dimensional propagation axis. This configuration has been extensively studied analytically~\cite{sheremetWaveguideQuantumElectrodynamics2022,royStronglyInteractingPhotons2017,rouletTwoPhotonsAtomic2016,greenbergSinglephotonScatteringQubit2023,domokosQuantumDescriptionLight2002,shenCoherentSinglePhoton2005,shenStronglycorrelatedMultiparticleTransport2007,shenStronglyCorrelatedTwoPhoton2007,fanInputOutputFormalismFewPhoton2010,pletyukhovScatteringMasslessParticles2012,chenPropagatorsJaynesCummingsModel2002,ralleyHongOuMandelInterferenceSingle2015,shiMultiphotonScatteringTheory2015,schneiderGreensfunctionFormalismWaveguide2016,nysteenScatteringTwoPhotons2015}. It therefore provides a valuable benchmark with a reliable analytical baseline against which the numerical simulations can be compared, allowing us to identify and correct truncation effects. Notice that, because the one-dimensional coupling has dimensions of the square root of an energy, the quantity $\gamma_{0}$ has dimensions of energy. Physically, it corresponds to the decay rate of the TLS into the photonic modes~\cite{shenCoherentSinglePhoton2005,rouletTwoPhotonsAtomic2016}. We note that this is a standard Markovian waveguide-QED setting for which efficient time-domain simulation methods already exist~\cite{bundgaard-nielsenWaveguideQEDjlEfficientFramework2025,regidorQwaveMPSEfficientOpensource2026}. Our purpose is not to compete with such methods in this regime, but rather to use a well-understood reference problem to quantify cutoff-induced deviations in frequency-truncated photon-matter simulations.

The second model is a two-dimensional atomic array, shown in the model-overview block of Fig.~\ref{fig:model_calibration_summary}, inspired by Refs.~\cite{havukainenQuantumSimulationsOptical1999,obaFastSimulationMultiphoton2024}. In this case, several TLSs are placed at fixed positions and coupled to a two-dimensional photonic continuum through position-dependent phase factors. This model is more challenging because no simple analytical baseline is available, making reliable and efficient numerical simulations particularly important. We show how the analysis developed for the waveguide benchmark can be extended to this setting. Since the two-dimensional couplings are dimensionless, the quantities $d_{0,j}$ are dimensionless as well.

\subsection{Roadmap of the results}

\begin{figure*}[t]
    \centering
    \includegraphics[width=\linewidth]{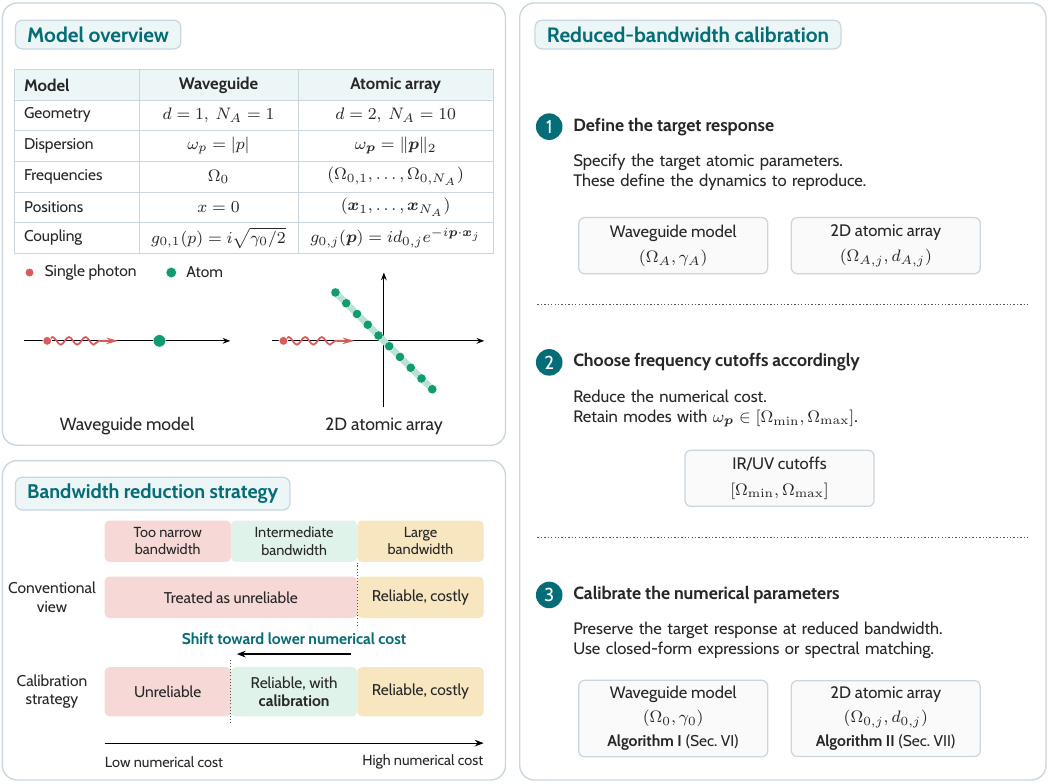}
    \caption{Overview of the two model classes, the bandwidth reduction strategy, and the corresponding reduced-bandwidth calibration procedures.}
    \label{fig:model_calibration_summary}
\end{figure*}
Before we enter the technical derivations, Fig.~\ref{fig:model_calibration_summary} summarizes the logic of the paper. The central objective is to reduce the simulation bandwidth, i.e., to retain only photonic modes whose frequencies lie in a finite window $[\ir,\uv]$, while preserving the physical response of the photon-matter device. Beyond being a practical necessity for finite numerical simulations, such frequency windows also have a natural physical interpretation. First, any qubit modality would have arguably narrow and specific operating frequencies. Furthermore, photonic quantum-network architectures are expected to rely on optical fibers to transport quantum states over long distances~\cite{kimbleQuantumInternet2008a,tanzilliPhotonicQuantumInformation2005}. In this context, propagation losses strongly favor specific wavelength bands, for example the telecom window around $1550\,\mathrm{nm}$~\cite{miyaUltimateLowlossSinglemode1979}. The photonic processes relevant to a given network component are therefore not expected to occupy the entire electromagnetic continuum, but rather a finite operational bandwidth around the carrier frequencies of interest. This makes reduced-bandwidth simulations not only computationally attractive, but also aligned with the way photonic devices are used in practice.

From a numerical perspective, a natural expectation would be that the window only needs to contain the photon frequencies and the atomic resonances relevant to the process being simulated. In Sec.~\ref{sec:bandwidth_diagnosis}, the waveguide benchmark shows that this expectation is not sufficient. Even in the single-photon sector, reducing the frequency window can change the effective resonance frequency and  decay rate seen in the simulated scattering response. This motivates the distinction, used throughout the paper, between target parameters and the numerical parameters supplied to the truncated Hamiltonian. We interpret these shifts as finite-bandwidth renormalization effects~\cite{peskinIntroductionQuantumField1995,collinsRenormalizationIntroductionRenormalization1984}. Rather than increasing the bandwidth until the shifts become negligible, we use the band-limited Green response to calibrate the numerical parameters directly. Figure~\ref{fig:model_calibration_summary} shows how this idea leads to an explicit prescription in the waveguide model and to a spectral-matching procedure in the two-dimensional array.

The scope of the present work is deliberately the single-photon sector. This does not address the full multiphoton scattering problem yet, but it fixes a necessary consistency condition for any simulation built from the same truncated Hamiltonian: if the one-photon response is already shifted by the bandwidth choice, multiphoton predictions inherit this error before genuinely nonlinear effects are even considered. Establishing and testing this one-excitation calibration is therefore a prerequisite for future extensions to multiphoton dynamics in quantum-network simulations.

\section{Construction of a finite dimensional Hamiltonian and its dynamics}
\label{sec:cutoff_regularization}

We now describe how a finite dimensional Hamiltonian is constructed for numerical simulations and examine how the dimension of the resulting numerical model scales with the number of retained photonic modes.

\subsection{Box discretization}

To begin with, the continuum must be regularized to be handled numerically. To do so, we place the system in a $d$-dimensional hypercube of side lengths $L_1, \dots, L_d$, hence volume $V = L_1 \dots L_d$. Imposing periodic boundary conditions discretizes the momentum variable as~\cite{obaFastSimulationMultiphoton2024,havukainenQuantumSimulationsOptical1999}
\begin{align*}
    \bm{p}_{\bm{n}} = \left(\Delta p_1 n_1, \dots, \Delta p_d n_d \right), \quad \bm{n}=(n_1,\dots,n_d)\in\mathbb{Z}^d,
\end{align*}
with $\Delta p_i = 2\pi / L_i$. Accordingly, continuum momentum integrals are replaced by discrete sums
via
\begin{align*}
    \int d^d\bm{p} \;\longrightarrow\; \frac{(2\pi)^d}{V} \sum_{\bm{n}}, \qquad \int \frac{d^d\bm{p}}{(2\pi)^d} \;\longrightarrow\; \frac{1}{V}\sum_{\bm{n}}.
\end{align*}
We denote by $\omega_{\bm{n}}=\omega_{\bm{p}_{\bm{n}}}$ the discrete mode frequencies. To obtain discrete bosonic operators satisfying standard commutation relations, while keeping the dimensions of the couplings consistent, we introduce
\begin{align*}
    a_{\bm{n}} = \frac{a(\bm{p}_{\bm{n}})}{\sqrt{V}}, \qquad g_{0,j\bm{n}} = \frac{g_{0,j}(\bm{p}_{\bm{n}})}{\sqrt{V}}.
\end{align*}
The discrete Hamiltonian is therefore expressed as
\begin{align*}
    \begin{split}
    H &= \sum_{\bm{n}} \omega_{\bm{n}} a_{\bm{n}}^{\dagger} a_{\bm{n}} + \sum_j \Omega_{0,j} \sigma_j^+ \sigma_j^- \\
    & \hspace*{5em}+ \sum_{j, \bm{n}} \Big[ g_{0,j\bm{n}} a_{\bm{n}}\sigma_j^+ + g_{0,j\bm{n}}^* a_{\bm{n}}^\dagger\sigma_j^- \Big].
    \end{split}
\end{align*}

\subsection{Frequency windows and retained modes}
To obtain a finite and computationally tractable Hilbert space, the discretized Hamiltonian must be restricted to a finite set of photonic modes~\cite{obaFastSimulationMultiphoton2024,woodsSimulatingBosonicBaths2015,arranzregidorModelingQuantumLightmatter2021}. We impose infrared (IR) and ultraviolet (UV) cutoffs on the photon energies, chosen such that the resulting frequency window contains the energy scales relevant to the simulated dynamics. The mode sum is therefore restricted to
\begin{align*}
    \mathcal{I}_{\ir,\uv} = \left\{ \bm{n}\in\mathbb{Z}^d \,\middle|\, \omega_{\bm{n}}\in[\ir,\uv] \right\},
\end{align*}
and the number of retained photonic modes is 
\begin{align*}
    N_{\mathrm{modes}} = \left|\mathcal{I}_{\ir,\uv}\right|.
\end{align*}
Both models considered in this work (see Fig.~\ref{fig:model_calibration_summary}) assume the vacuum dispersion relation for the propagating photons. Accordingly, in units where $c=1$, the discrete mode frequencies are given by $\omega_{\bm{n}}=\Delta p \left\lVert \bm{n} \right\rVert_2$. The retained index set therefore reads
\begin{align*}
    \mathcal{I}_{\ir, \uv} = \left\{ \bm{n} \in \mathbb{Z}^d \, \middle| \, \frac{\ir}{\Delta p} \leq \left\lVert \bm{n} \right\rVert_2 \leq \frac{\uv}{\Delta p} \right\}.
\end{align*}
Hence, the number of modes is approximately
\begin{align*}
    \begin{split}
    N_{\mathrm{modes}} &\simeq \mathrm{Vol}\!\left(B_d\left(\frac{\uv}{\Delta p}\right)\right) - \mathrm{Vol}\!\left(B_d\left(\frac{\ir}{\Delta p}\right)\right) \\
    &= \frac{\mathcal{V}_d}{(\Delta p)^d} \left(\uv^d-\ir^d\right),
    \end{split}
\end{align*}
where $B_d(R)$ is the $d$-dimensional ball of radius $R$ and $\mathcal{V}_d=\pi^{d/2}/\Gamma(d/2+1)$ is the volume of the unit $d$-ball. Additional polarization or channel degeneracies would simply multiply this count~\cite{obaFastSimulationMultiphoton2024}. The numerical bandwidth $|\uv-\ir|$ strongly affects the number of modes retained in the simulation. As we will see, this dependence has direct practical implications for computational efficiency.

\subsection{Numerical Hilbert space dimensionality}

Let us now comment on how the dimensionality of the numerical Hilbert space scales with the number of retained modes. The fixed-excitation construction introduced above allows us to restrict the dynamics to a sector with a finite number of excitations $N$. For a given number $q$ of excited atoms, the photonic part contains $N-q$ photons and is spanned by states of the form
\begin{align*}
    \ket{\bm{p}_{\bm{n}_1},\dots,\bm{p}_{\bm{n}_{N-q}}, e_{j_1}, \dots, e_{j_q}},
\end{align*}
with $\bm{n}_\ell\in\mathcal{I}_{\ir,\uv}$. The number of photonic amplitudes in the sector containing $q$ atomic excitations therefore scales as\footnote{Because photons obey bosonic symmetry, permutations of the momentum labels represent the same physical state. For the present scaling argument, accounting for this redundancy only changes a combinatorial prefactor at fixed $N$ and therefore does not affect the scaling.}
\begin{align*}
    \mathcal{O}\left(N_{\mathrm{modes}}^{N-q}\right).
\end{align*}
Since, for fixed $N_A$, the number of atomic configurations is independent of $N_{\mathrm{modes}}$, the leading contribution arises from the purely photonic sector, corresponding to $q=0$. Hence, the Hilbert-space dimension of a simulation involving $N$ excitations scales as
\begin{align*}
    \dim\left(\mathcal{H}_{\mathrm{num}}\right) = \mathcal{O}\Bigg( \left( \frac{\uv^d-\ir^d}{(\Delta p)^d} \right)^N \Bigg).
\end{align*}
This explicit dependence on the retained bandwidth is the main computational reason for narrowing the frequency window whenever this can be done without changing the physical response of the simulated model.

\subsection{Runge-Kutta-based propagation}
\label{subsec:rk_simulation}

It is useful to remove the free evolution generated by $H_{\mathrm{free}}$ in order to focus solely on the nontrivial interaction term in the upcoming numerical simulations~\cite{havukainenQuantumSimulationsOptical1999}. We therefore work in the interaction picture with respect to $H_{\mathrm{free}}$, where the state then evolves according to
\begin{align*}
    \partial_t\ket{\psi(t)} = -i V(t)\ket{\psi(t)}, \quad V(t)=e^{iH_{\mathrm{free}}t}H_Ie^{-iH_{\mathrm{free}}t}.
\end{align*}
Using the free evolution of the photon and TLS operators, this gives
\begin{align}
    \begin{split}
    V(t) & = \sum_{j, \bm{n}} \Big[ g_{0,j\bm{n}}\, e^{-i(\omega_{\bm{n}}-\Omega_{0,j})t} a_{\bm{n}}\,\sigma_j^+ \\
    & \hspace*{6em} + g_{0,j\bm{n}}^*\, e^{i(\omega_{\bm{n}}-\Omega_{0,j})t} a_{\bm{n}}^\dagger\,\sigma_j^- \Big].
    \end{split}
    \label{eq:interacting_hamilt}
\end{align}
Starting from an initial state $\ket{\psi_0}$, the simulation time window $[0,T]$ is discretized into time steps
\begin{align*}
    t_i = i\Delta t, \qquad t_0 = 0, \qquad t_{N_{\mathrm{RK4}}} = T,
\end{align*}
with $N_{\mathrm{RK4}} \sim T/\Delta t$. Following Ref.~\cite{havukainenQuantumSimulationsOptical1999}, the time evolution is numerically integrated using a fourth-order Runge--Kutta (RK4) scheme. A sequence of states $\ket{\psi_i}=\ket{\psi(t_i)}$ is iteratively generated according to
\begin{align}
    \begin{split}
        \ket{\psi_{i+1}} &= \ket{\psi_i} + \ket{\delta \psi_i}, \\
        \ket{\delta \psi_i} & = \frac{\ket{\delta\psi_{i1}} + 2\ket{\delta\psi_{i2}} + 2\ket{\delta\psi_{i3}} + \ket{\delta\psi_{i4}}}{6},
    \end{split}
    \label{eq:rk4_propagation}
\end{align}
where the intermediate RK4 increments are given by~\cite{havukainenQuantumSimulationsOptical1999}
\begin{align}
    \begin{dcases}
    \ket{\delta \psi_{i1}} &= -i\Delta t\,V(t_i)\ket{\psi_i}\\
    \ket{\delta \psi_{i2}} &= -i\Delta t\,V\!\left(t_i+\frac{\Delta t}{2}\right) \left( \ket{\psi_i} + \frac{1}{2}\ket{\delta\psi_{i1}} \right)\\
    \ket{\delta \psi_{i3}} &= -i\Delta t\,V\!\left(t_i+\frac{\Delta t}{2}\right) \left( \ket{\psi_i} + \frac{1}{2}\ket{\delta\psi_{i2}} \right)\\
    \ket{\delta \psi_{i4}} &= -i\Delta t\,V(t_i+\Delta t) \left( \ket{\psi_i} + \ket{\delta\psi_{i3}} \right)
    \end{dcases}.
    \label{eq:rk4_increment_states}
\end{align}
The computational cost of a single RK4 iteration is dominated by the evaluation of these intermediate states. Appendix~\ref{appendix:detailed_rk_scheme} details how the scheme is vectorized to compute the corresponding increments and shows that their evaluation costs at worst $\mathcal{O}\left(\dim\mathcal{H}_{\mathrm{num}}\right)$. This leads to an overall propagation cost
\begin{align*}
    \mathcal{O} \Bigg( N_{\mathrm{RK4}} \times \left( \frac{\uv^d-\ir^d}{(\Delta p)^d} \right)^N \Bigg).
\end{align*}
Reducing the simulated bandwidth $\uv-\ir$ is therefore highly desirable from a computational perspective. However, reducing it too aggressively may compromise the physical reliability of the simulation. This creates a tradeoff between computational cost and physical accuracy, which must be controlled if reduced-bandwidth simulations are to be used reliably. The remainder of this work investigates this tradeoff, starting from a minimal benchmark model derived from the generic framework introduced above.

\section{Assessing bandwidth effects with a single photon}

\label{sec:bandwidth_diagnosis}

In this section, we first focus on the waveguide model to diagnose finite-bandwidth effects. We define the single-photon reflection experiment and its analytical reference response, before using it to identify nontrivial signatures of IR and UV truncation.
\subsection{Waveguide scattering experiment}

Recall the specifications of the waveguide model:
\begin{align*}
    N_A = 1, \qquad d = 1, \qquad \omega_p = |p|, \qquad g_{0}(p) = i\sqrt{\gamma_{0}/2}.
\end{align*}
In the simulation considered in this section, we prepare a monochromatic right-propagating single photon incident from the region $x<0$ and send it toward the TLS to evaluate the simulated scattering response. Restricting the dynamics to the single-excitation sector is deliberate: the analytical response is well understood, so any deviation from the known scattering behavior can be attributed directly to the finite momentum-space window.

For an incoming right-propagating photon with momentum $p>0$ and energy $\omega=\omega_p=|p|$, energy conservation restricts the outgoing photon to modes with momenta $\pm p$. In particular, for a single TLS with target transition frequency $\Omega_{A}$ and target decay rate $\gamma_{A}$, the theoretical reflection probability is given by~\cite{shenCoherentSinglePhoton2005,rouletTwoPhotonsAtomic2016,royStronglyInteractingPhotons2017,fanInputOutputFormalismFewPhoton2010}
\begin{align}
    \mathcal{R}^{(\mathrm{th})}(\omega) = |r_\omega|^2, \qquad r_{\omega} = \frac{-i\gamma_{A}/2} {\omega-\Omega_{A}+i\gamma_{A}/2},
    \label{eq:single_photon_baseline_amplitudes}
\end{align}
where $r_\omega$ denotes the corresponding reflection amplitude. We will investigate how accurately the numerical simulation reproduces this response.

\subsection{Wave-packet initialization}

In this simulation, we choose arbitrary target values $(\Omega_{A},\gamma_{A})$ for the TLS parameters and configure the Hamiltonian using numerical parameters that coincide with these target values:
\begin{align*}
\Omega_{0} = \Omega_{A}, \qquad \gamma_{0} = \gamma_{A}.
\end{align*}
This identification is made independently of the remaining simulation parameters. The momentum grid is defined by $p_n=n\Delta p$. Since the photon dispersion relation is $\omega_p=|p|$, the numerical window imposed on the photon energies is therefore
\begin{align*}
\mathcal{I}_{\ir,\uv} = \left\{ n \in \mathbb{Z} \,\middle|\, \ir \leq |n|\Delta p \leq \uv \right\}.
\end{align*}
The sign of $n$ then labels the propagation direction: positive and negative momenta correspond to right- and left-propagating photons, respectively. To emulate the monochromatic response of Eq.~\eqref{eq:single_photon_baseline_amplitudes}, the incoming state is prepared as a right-propagating Gaussian wave packet that is narrow in frequency. It has central momentum $p_{\mathrm{init}}>0$, momentum-space width $\sigma_p$, and initial real-space center $x_{\mathrm{init}}$, while the TLS is prepared in its ground state:
\begin{align*}
    \ket{\psi_0} = \sum_{\substack{n\in\mathcal{I}_{\ir,\uv}\\ n>0}} c_{n,0}\, \ket{p_n},
\end{align*}
with coefficients given by~\cite{havukainenQuantumSimulationsOptical1999,obaFastSimulationMultiphoton2024}
\begin{align*}
    c_{n,0} \propto \exp\!\left[ -\frac{(p_n-p_{\mathrm{init}})^2}{4\sigma_p^2} - ip_n x_{\mathrm{init}} \right],
\end{align*}
where the proportionality constant is fixed numerically to ensure
normalization within the finite momentum grid retained by the energy window $[\ir,\uv]$. Choosing $\sigma_p\ll \gamma_{A}$ allows one to probe the monochromatic limit~\cite{rouletTwoPhotonsAtomic2016}. Starting from this initial configuration, the RK4 propagation scheme generates the list of states
\begin{align*}
    \ket{\psi_i} = \sum_{n\in\mathcal{I}_{\ir,\uv}} c_{n,i}\, \ket{p_n} + b_i\ket{e}.
\end{align*}
After propagation, reflection corresponds to probability transferred to the negative-momentum sector. We therefore compute
\begin{align*}
    \mathcal{R} = \sum_{\substack{n\in\mathcal{I}_{\ir,\uv}\\ n<0}} |c_{n,\, N_{\text{step}}}|^2.
\end{align*}
The numerical reflection probability can then be directly compared with the analytical prediction evaluated at $\omega_{\mathrm{init}} = p_{\mathrm{init}}$. The simulation parameters used in this benchmark are reported in Table~\ref{tab:simulation_parameters}. The RK4 time step is chosen sufficiently small so that the numerical integration error is negligible on the scale of the bandwidth effects studied here; its contribution to the computational cost is discussed later.

\begin{table}[h]
\centering
\begin{tabular}{|l l|}
\hline
\multicolumn{2}{|c|}{\textbf{Target TLS parameters}} \\
\hline
Atomic frequency $\Omega_{A}$ &  $10\pi$ \\
Target decay rate $\gamma_{A}$ & $\pi$ \\
\hline
\multicolumn{2}{|c|}{\textbf{Photon parameters}} \\
\hline
Central momentum $p_{\mathrm{init}}$ & variable \\
Associated energy $\omega_{\mathrm{init}}$ & $\displaystyle [\Omega_{A}-\frac{\gamma_{A}}{2} ,\,\Omega_{A}+\frac{\gamma_{A}}{2}]$ \\
Momentum spread $\sigma_p$ & $0.05\pi$ \\
Initial position $x_{\mathrm{init}}$ & $-25$ \\
\hline
\multicolumn{2}{|c|}{\textbf{Numerical simulation parameters}} \\
\hline
Waveguide length $L$ & $100$ \\
Simulation time $T$ & $50$ \\
RK4 time step $\Delta t$ & $0.01$ \\
Numerical TLS frequency $\Omega_{0}$ & $10\pi$ \\
Numerical  decay rate $\gamma_{0}$ & $\pi$ \\
\hline

\end{tabular}
\vspace*{1em}
\caption{Parameters used in the waveguide benchmark simulations}
\label{tab:simulation_parameters}
\end{table}

\subsection{Cutoff-induced parameter shifts}

\begin{figure*}[t]
    \centering
    \includegraphics[width=\linewidth]{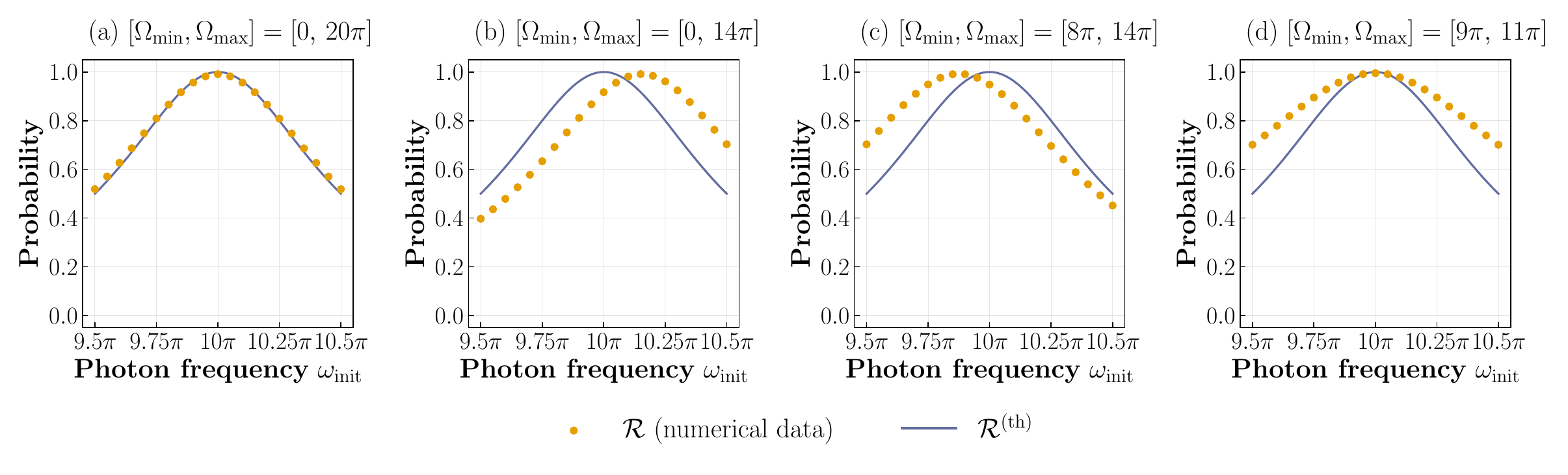}
    \caption{Comparison between the numerically obtained reflection coefficient $\mathcal{R}$ and the prediction $\mathcal{R}^{\text{(th)}}$ across various frequency windows.}
   \label{fig:reflection_multi_cutoff_experiment_uncorrected}
\end{figure*}

To characterize finite-bandwidth effects, we analyze the system response for fixed frequency windows chosen to include the relevant energy scales. The underlying intuition is that an accurate simulation should primarily require resolving the frequency modes involved in the scattering process, while modes far from the atomic resonance are expected to contribute negligibly. Specifically, we reconstruct the reflection profile $\mathcal{R}$ by performing scattering simulations for several central momenta $p_{\mathrm{init}}>0$, with $\omega_{\mathrm{init}} = p_{\mathrm{init}}$ sampled in the interval $[\Omega_{A}-\gamma_{A}/2,\,\Omega_{A}+\gamma_{A}/2] = [9.5\pi, 10.5\pi]$. Following this intuition, one might expect that choosing an IR cutoff $\ir$ close to $9\pi$ and a UV cutoff $\uv$ close to $11\pi$ would be sufficient to reproduce the theoretical profile $\mathcal{R}^{(\mathrm{th})}$ accurately. We then compare the trends of $\mathcal{R}$ and $\mathcal{R}^{(\mathrm{th})}$ using first a deliberately oversized frequency window, before progressively reducing the bandwidth. For this purpose, we consider the following four frequency windows:
\begin{align*}
    [\ir, \uv] = [0, 20\pi], [0, 14\pi], [8\pi, 14\pi], [9\pi, 11\pi].
\end{align*}
The first and last windows are deliberately chosen to be symmetric around the atomic frequency $\Omega_A$, while the two intermediate windows are selected as representative steps in the progressive bandwidth reduction.

The corresponding results are shown in the subpanels of Fig.~\ref{fig:reflection_multi_cutoff_experiment_uncorrected}. Surprisingly, only the first frequency window yields satisfactory agreement. Across the four windows, the overall functional form of the reflection profile is reproduced. However, as the bandwidth is reduced, both the location of the reflection peak, expected to occur at resonance, and the width of the curve are modified. This suggests that the effective parameters governing the simulated dynamics no longer match the target TLS parameters $\Omega_{A}$ and $\gamma_{A}$, despite choosing the Hamiltonian parameters as $(\Omega_{0},\gamma_{0})=(\Omega_{A},\gamma_{A})$.

In subpanels (a) and (d) of Fig.~\ref{fig:reflection_multi_cutoff_experiment_uncorrected}, the target TLS frequency $\Omega_{A}=10\pi$ lies exactly at the center of the numerical frequency window. For these two frequency windows, the peak location is accurately predicted, suggesting that centering the numerical frequency window around $\Omega_{A}$ makes the Hamiltonian parameterization $\Omega_{0} = \Omega_{A}$ consistent. However, the numerical reflection profile broadens as the window is narrowed in subpanel (d), indicating that the simulated  decay rate does not match the target decay rate $\gamma_{A}$.

These simulations suggest that even when the frequency window covers the relevant energy range, an intermediate regime remains in which the numerical reflection profile has the correct qualitative shape, but is governed by shifted effective parameters. In this regime, the numerical parameters $(\Omega_{0},\gamma_{0})$ do not generate scattering dynamics corresponding to their nominal values. This motivates explicitly distinguishing the numerical Hamiltonian parameters $(\Omega_{0},\gamma_{0})$ from the target parameters $(\Omega_{A},\gamma_{A})$.

\subsection{Qualitative interpretation and calibration strategy}

\begin{figure}
    \centering
    \includegraphics[width=\linewidth]{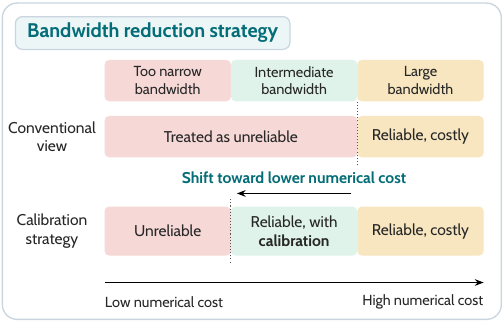}
    \caption{Illustration of the computational benefits of a calibration procedure that makes an intermediate-bandwidth regime accessible.}
    \label{fig:bandwidth_reduc}
\end{figure}

Coupling a quantum system to a continuum of modes generally modifies its effective parameters, giving rise, for instance, to shifts in resonance frequencies and decay rates. Such effects, closely related to the well-known Lamb shift~\cite{lambFineStructureHydrogen1947}, are naturally described within the framework of renormalization~\cite{peskinIntroductionQuantumField1995,collinsRenormalizationIntroductionRenormalization1984}. In this context, the parameters appearing in the Hamiltonian do not necessarily coincide with those governing the observable scattering response. This provides a natural explanation for the impact of frequency cutoffs on the dynamics: restricting the continuum to a finite simulation bandwidth modifies the effective parameters, which we interpret as a renormalization effect. Related effects have been discussed in several photon-scattering models involving TLSs~\cite{malekakhlaghCutofffreeCircuitQuantum2017,sheremetWaveguideQuantumElectrodynamics2022,martinezMixedPerturbativenonperturbativeTreatment2024,greenbergSinglephotonScatteringQubit2023}. However, to the best of our knowledge, they have not yet been exploited to reduce the numerical cost of photon-matter simulations.

Rather than increasing the numerical bandwidth until the simulation reaches the desired accuracy, one can instead work with a narrower frequency window and recalibrate the numerical parameters $(\Omega_{0},\gamma_{0})$ to recover the target parameters $(\Omega_{A},\gamma_{A})$. This strategy, introduced earlier in the paper and further illustrated in Fig.~\ref{fig:bandwidth_reduc}, enables accurate scattering simulations with a reduced number of frequency modes, thereby lowering the overall computational cost.

In the remainder of this paper, we set up the theoretical framework underlying this recalibration strategy for the class of photon-matter Hamiltonians introduced above. We then illustrate its practical implementation on the one-dimensional toy-model testbed considered in the previous section, and show that the same target scattering response can be recovered with substantially smaller bandwidths.

\section{Single-photon response from the Green matrix}
\label{sec:green_matrix_response}

This section develops the theoretical framework underlying the calibration of reduced-bandwidth simulations at the single-photon level. Our analysis relies primarily on Green's functions and related techniques, which are standard tools for studying photon scattering by quantum matter systems~\cite{seeDiagrammaticApproachMultiphoton2017,stefanoFeynmandiagramsApproachQuantum2017,sheremetWaveguideQuantumElectrodynamics2022}. 

\subsection{Main theorem}

For compactness in the general Green-matrix derivation below, we write the model parameters $\Omega_j=\Omega_{0,j}$ and $g_j=g_{0,j}$. The following theorem allows one to precisely isolate the origin of apparent truncation-dependent behaviors in single-photon scattering simulations.

\begin{theorem}[Effective scattering amplitude]
\label{thm:green_operator}

Consider the Hamiltonian introduced in Sec.~\ref{sec:hamiltonian_model}, truncated over a finite frequency window $[\ir,\uv]$, and an incoming photon with momentum $\bm{p}$ and energy $\omega=\omega_{\bm{p}}$. In the truncated model, the probability of observing a photon scattered into a momentum mode $\bm{q}$ is governed by the scattering amplitude
\begin{align}
    \begin{split}
    &\mathcal{T}_{\ir,\uv}(\bm{q},\bm{p},\omega, \{\bm{\Omega}\}, \{\bm{g}\}) \\
    & \hspace*{7em} = -i\,\bm{g}^{\dagger}(\bm{q}) \mathsf{G}_{\mathrm{eff}}(\omega+i0^+) \bm{g}(\bm{p}),
    \end{split}
    \label{eq:single_photon_transition_amplitude}
\end{align}
where energy conservation restricts the outgoing momentum to $\omega_{\bm{q}}=\omega_{\bm{p}} = \omega$. The relevant mathematical objects entering this expression are defined below.
\begin{proof}
    See Subsections~\ref{subsec:green_operator}, \ref{subsec:band_limited_self_energy}.
\end{proof}
\end{theorem}
The effective scattering amplitude in Eq.~\eqref{eq:single_photon_transition_amplitude} involves the following quantities:
\begin{itemize}
\item The coupling vector associated with a photon mode is defined as
\begin{align*}
\bm{g}(\bm{p})
=
\begin{pmatrix}
g_1(\bm{p}), \dots, g_{N_A}(\bm{p})
\end{pmatrix}^{\mathsf{T}}.
\end{align*}

\item The cutoff-dependent atomic Green matrix, formed from the elements $G_{jk}(z)=\bra{e_j}G(z)\ket{e_k}$ of the Green operator $G(z)=(z-H)^{-1}$, is given by
\begin{align*}
    \mathsf{G}_{\mathrm{eff}}^{(\ir,\uv)}(z)
    =
    \left[
        z\mathsf{I}_{N_A}
        - \mathsf{\Omega}
        - \mathsf{\Sigma}_{\mathrm{eff}}^{(\ir,\uv)}(z)
    \right]^{-1},
\end{align*}
where $\mathsf{I}_{N_A}$ is the $N_A\times N_A$ identity matrix and $\mathsf{\Omega} = \operatorname{Diag}(\bm{\Omega})$, with $\bm{\Omega}=(\Omega_1,\ldots,\Omega_{N_A})^{\mathsf{T}}$.

\item The band-limited self-energy matrix entering the resolvent reads
\begin{align}
    \mathsf{\Sigma}^{(\ir, \uv)}_{\mathrm{eff}}(z) = \int_{\ir}^{\uv} d\omega' \frac{\mathsf{J}(\omega')}{z-\omega'}.
\end{align}
Here, $\mathsf{J}$ is the spectral density matrix previously defined in Eq.~\eqref{eq:spectral_density_matrix}. Consequently, all finite-bandwidth effects enter the scattering amplitude through this effective self-energy.

\end{itemize}

\subsection{Scattering matrix and projected resolvent}
\label{subsec:green_operator}
We consider a scattering scenario in which an incoming photon propagates towards the device, interacts with it, and subsequently propagates away in an outgoing photonic mode. To characterize this process, we follow the standard scattering-theory framework~\cite{peskinIntroductionQuantumField1995}, in which the scattering operator $S$ relates incoming free photonic states to outgoing free photonic states. In the single-photon sector, the relevant quantities are its matrix elements $\langle \bm{q} | S | \bm{p} \rangle$, which encode the probability amplitudes for an incoming photon with momentum $\bm{p}$ to be scattered into an outgoing mode with momentum $\bm{q}$. These matrix elements can be expressed as~\cite{peskinIntroductionQuantumField1995,cohen-tannoudjiPhotonsAtomsIntroduction1989,pletyukhovScatteringMasslessParticles2012}
\begin{align*}
    \begin{split}
    \langle \bm{q} | S | \bm{p} \rangle &= (2\pi)^d \delta^{(d)}(\bm{q}-\bm{p}) \\
    & \hspace*{1cm} - 2i\pi \delta(\omega_{\bm{q}}-\omega_{\bm{p}}) \langle \bm{q} | T(\omega+i0^+) | \bm{p} \rangle .
    \end{split}
\end{align*}
Here, the nontrivial part of the $S$-matrix is encoded in the transition operator $T(z)$, commonly referred to as the $T$-operator in scattering theory~\cite{peskinIntroductionQuantumField1995}. Its matrix elements are evaluated at $z=\omega+i0^+$ with $\omega=\omega_{\bm{p}}=\omega_{\bm{q}}$, because of energy conservation. Accordingly, we define the amplitude
\begin{align}
    \mathcal{T}_{\ir, \uv}(\bm{q},\bm{p},\omega, \{\bm{\Omega}\}, \{\bm{g}\}) = -i \bra{\bm{q}} T(\omega + i0^+) \ket{\bm{p}},
\end{align}
which fully determines the scattering response of the system in the single-photon sector. In particular, the operator $T(z)$ satisfies~\cite{pletyukhovScatteringMasslessParticles2012,shiMultiphotonScatteringTheory2015}
\begin{align}
    T(z) = H_I + H_I G(z) H_I,
    \label{eq:T_matrix_green_operator}
\end{align}
where $G(z)$ is the \emph{Green operator}, or resolvent. It is defined for $\operatorname{Im} z>0$, as
\begin{align*}
    G(z) = \Big[z - H \Big]^{-1} \equiv \frac{1}{z-H}.
\end{align*}
Since the interaction Hamiltonian $H_I$ does not directly couple one-photon ground-state configurations to each other under the RWA, the first term in Eq.~\eqref{eq:T_matrix_green_operator} does not contribute to matrix elements between one-photon scattering states. Inserting complete sets of atomic states gives
\begin{align*}
    \begin{split}
    \bra{\bm{q}}T(z)\ket{\bm{p}} = \sum_{j,k=1}^{N_A} \bra{\bm{q}} H_I \ket{e_j} \bra{e_j}G(z)\ket{e_k} \bra{e_k} H_I \ket{\bm{p}},
    \end{split}
\end{align*}
and using the explicit matrix elements
\begin{align*}
    \bra{\bm{q}} H_I \ket{e_j}=g_j^*(\bm{q}), \qquad \bra{e_k} H_I \ket{\bm{p}}=g_k(\bm{p}),
\end{align*}
one obtains
\begin{align*}
    \bra{\bm{q}}T(z)\ket{\bm{p}}
    =
    \bm{g}(\bm{q})^{\dagger}
    \mathsf{G}(z)
    \bm{g}(\bm{p}).
\end{align*}
Evaluating this expression with $z=\omega+i0^+$ and $\omega_{\bm{p}}=\omega_{\bm{q}}=\omega$, and including the remaining factor $-i$ gives Eq.~\eqref{eq:single_photon_transition_amplitude}. Thus, the nontrivial single-photon scattering response is entirely encoded in the projected atomic Green matrix and in the external coupling vectors. 

\subsection{Band-limited self-energy}
\label{subsec:band_limited_self_energy}

Let us now express the atomic Green matrix $\mathsf{G}(z)$ as a function of the self-energy. For each atomic basis state $\ket{e_k}$, define
\begin{align*}
    \ket{\psi_k(z)}=G(z)\ket{e_k}, \qquad G_{jk}(z)=\braket{e_j|\psi_k(z)}.
\end{align*}
Accordingly, expanding $\ket{\psi_k(z)}$ in the single-excitation sector gives
\begin{align*}
    \ket{\psi_k(z)} = \int \frac{d^d\bm{p}}{(2\pi)^d} c_k(\bm{p},z)\ket{\bm{p}} + \sum_{j=1}^{N_A} G_{jk}(z)\ket{e_j}.
\end{align*}
By definition, $(z-H)\ket{\psi_k(z)}=\ket{e_k}$. Projection onto the one-photon and atomic basis states yields
\begin{align*}
    \begin{dcases}
        (z-\omega_{\bm{p}})c_k(\bm{p},z) - \sum_{m=1}^{N_A} g_m^*(\bm{p})G_{mk}(z) &= 0 \\
        (z-\Omega_j)G_{jk}(z) - \int \frac{d^d\bm{p}}{(2\pi)^d} g_j(\bm{p})c_k(\bm{p},z) &= \delta_{jk}
    \end{dcases}.
\end{align*}
The first equation gives
\begin{align*}
    c_k(\bm{p},z) = \sum_{m=1}^{N_A} \frac{ g_m^*(\bm{p})G_{mk}(z) }{ z-\omega_{\bm{p}} }.
\end{align*}
Substituting this expression into the second equation yields
\begin{align}
    \sum_{m=1}^{N_A} \left[ (z-\Omega_j)\delta_{jm} - \Sigma_{jm}(z) \right] G_{mk}(z) = \delta_{jk},
    \label{eq:relation_self_energy_and_green_indices}
\end{align}
with $\Sigma_{jm}(z)$ being the \emph{self-energy}, given by
\begin{align*}
    \Sigma_{jm}(z) = \int \frac{d^d\bm{p}}{(2\pi)^d} \frac{g_j(\bm{p})g_m^*(\bm{p})}{z-\omega_{\bm{p}}}.
\end{align*}
Eq.~\eqref{eq:relation_self_energy_and_green_indices} gives the matrix relation between the atomic Green matrix $\mathsf{G}$ and the self-energy $\mathsf{\Sigma}$ introduced in Theorem~\ref{thm:green_operator}. Finally, to isolate the role of the finite frequency window, we now make the self-energy contribution explicit. Using the spectral-density matrix $\mathsf{J}(\omega)$ defined in Eq.~\eqref{eq:spectral_density_matrix}, one obtains
\begin{align}
    \mathsf{\Sigma}(z) = \int d\omega' \frac{\mathsf{J}(\omega')}{z-\omega'}.
    \label{eq:general_self_energy}
\end{align}
Restricting the above integral to the integration domain $[\ir, \uv]$ gives the band-limited self-energy as stated in the theorem.

\subsection{Matching condition}

The calibration theorem (Theorem~\ref{thm:green_operator}) provides a practical means of restoring consistency in frequency-truncated photon-matter simulations in the single-photon sector. The truncated model should be calibrated by requiring its effective amplitude to reproduce a prescribed target response over the operational bandwidth:
\begin{align*}
    \begin{split}
    & \mathcal{T}_{\ir,\uv}(\bm{q},\bm{p};\omega, \{\bm{\Omega}\}, \{\bm{g}\}) \simeq f_{\mathrm{target}}(\bm{q},\bm{p},\omega), \\
    &
    \begin{dcases}
    &\forall \omega \in [\ir, \uv]\\
    &\omega_{\bm{p}}=\omega_{\bm{q}}=\omega     
    \end{dcases}.
    \end{split}
\end{align*}
Here, $f_{\mathrm{target}}$ denotes either an analytical target response or a reference response obtained using a wider frequency window. The notation retains the possible angular or modal dependence carried by $\bm{p}$ and $\bm{q}$, but these momenta are not independent variables: they are always evaluated at the same conserved energy $\omega$. We now illustrate how this condition is implemented in practice, first by returning to the waveguide benchmark and then by considering the two-dimensional geometry introduced above.

\section{Calibrated waveguide simulations}
\label{sec:waveguide_calibration}

In this section, we return to the toy one-dimensional model
\begin{align*}
N_A = 1, \quad d = 1, \quad \omega_p = |p|, \quad g_{0}(p) = i\sqrt{\gamma_{0}/2},
\end{align*}
to show a practical application of Theorem~\ref{thm:green_operator}. Thus, reduced-bandwidth simulations can remain accurate when consistent corrections are applied, supporting the strategy presented in Fig.~\ref{fig:model_calibration_summary}.

\subsection{Effective dynamics of the truncated model}

\begin{figure*}[t]
    \centering
    \includegraphics[width=\linewidth]{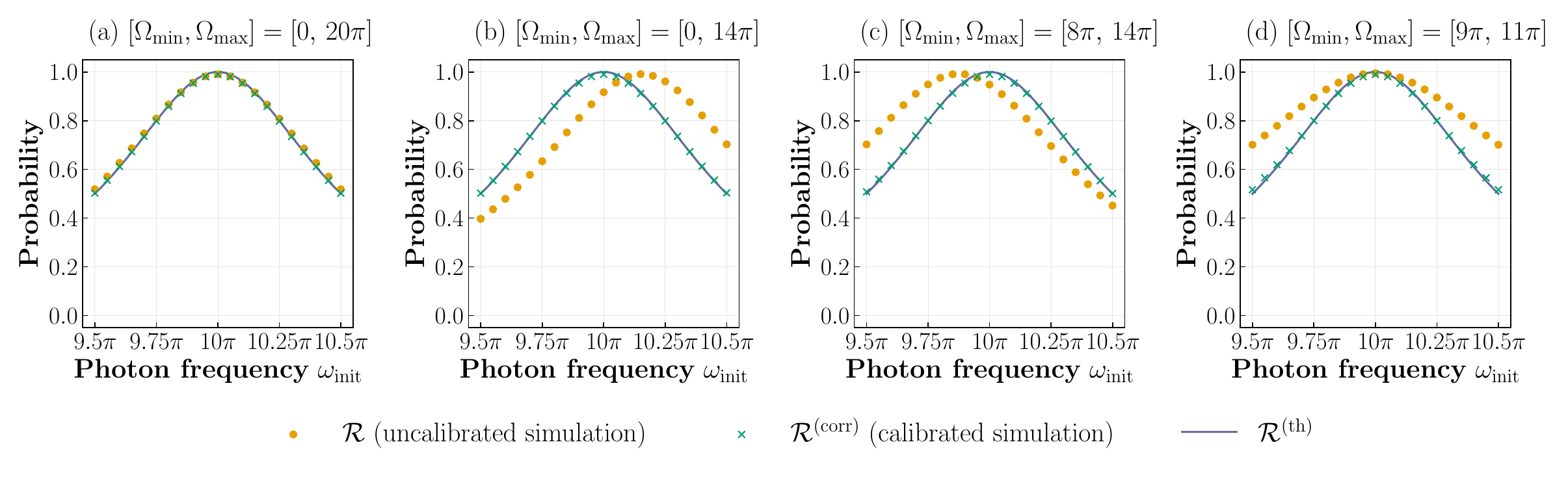}
    \caption{Comparison between the numerically obtained reflection coefficient $\mathcal{R}$ and the adjusted prediction $\mathcal{R}^{(\mathrm{corr})}$ derived from the target parameters, evaluated across different frequency windows. The uncalibrated prediction $\mathcal{R}^{(0)}$ is shown for reference.}
   \label{fig:reflection_multi_cutoff_experiment_corrected}
\end{figure*}

\begin{corollary}[Waveguide benchmark calibration]
\label{cor:renormalization_relations}
We now specialize Theorem~\ref{thm:green_operator} to the waveguide setup used in the benchmark model. For a right-propagating incoming photon of frequency $\omega$, the incoming and reflected momenta satisfy $p=\omega$ and $q=-\omega$, respectively. The dependence of the scattering amplitude on $(p,q,\omega)$ therefore reduces to a dependence on $\omega$ alone. With this simplified notation, the effective reflection amplitude emulated by the simulation reads
\begin{align*}
    \mathcal{T}_{\ir,\uv}(\omega,\Omega_0,\gamma_0)
    = r^{\mathrm{eff}}_{\omega}.
\end{align*}
Provided that the cutoffs $\ir,\uv$ are sufficiently detuned from the numerical TLS frequency $\Omega_0$, the effective reflection amplitude takes the Lorentzian form
\begin{align*}
    r_{\omega}^{\mathrm{eff}}
    = -\frac{i\gamma_{\mathrm{eff}}/2}
    {\omega-\Omega_{\mathrm{eff}}+i\gamma_{\mathrm{eff}}/2},
\end{align*}
where the effective parameters are given by
\begin{subequations}
\begin{align*}
    \Omega_{\mathrm{eff}}
    &= \Omega_0 - \frac{\gamma_0}{2\pi}
    f(\Omega_0,\gamma_0)
    \log\!\left(\frac{\uv-\Omega_0}{\Omega_0-\ir}\right),\\
    \gamma_{\mathrm{eff}}
    &= \gamma_0 f(\Omega_0,\gamma_0),
\end{align*}
\end{subequations}
with
\begin{align*}
    \begin{split}
        &f(\Omega_0,\gamma_0)\\
        &\quad= \Bigg[1-\frac{\gamma_0}{2\pi}
        \left(\frac{1}{\Omega_0-\ir}
        +\frac{1}{\uv-\Omega_0}\right)\Bigg]^{-1}.
    \end{split}
\end{align*}

\begin{proof}
As shown in Appendix~\ref{appendix:proof_crly1}, the exact reflection amplitude of the truncated model reads
\begin{align*}
    r_{\omega}^{\mathrm{eff}}
    = \frac{-i\gamma_0/2}{
    \displaystyle \omega-\Omega_0
    -\frac{\gamma_0}{2\pi}
    \log\frac{\omega-\ir}{\uv-\omega}
    +i\gamma_0/2}.
\end{align*}
Expanding its denominator to first order around $\omega=\Omega_0$ recovers the Lorentzian form and effective parameters stated above. The cutoff-detuning assumption ensures the validity of this approximation, corresponding to the green region of Fig.~\ref{fig:bandwidth_reduc}. The detailed derivation is provided in Appendix~\ref{appendix:proof_crly1}.
\end{proof}
\end{corollary}

This corollary supports the conclusion that one can work with a reduced frequency window chosen to contain the relevant frequencies of the problem, and recover the target response by imposing 
\begin{align}
    \Omega_{\mathrm{eff}}(\Omega_{0}, \gamma_{0}) = \Omega_{A}, \qquad \gamma_{\mathrm{eff}}(\Omega_{0}, \gamma_{0}) = \gamma_{A}.
    \label{eq:renorm_relation_waveguide}
\end{align}
For fixed user-chosen cutoffs $\ir$ and $\uv$, this calibration procedure is straightforward numerically, because the first relation can be rewritten as a scalar self-consistent equation for $\Omega_{0}$:
\begin{align}
    \Omega_{0} = \Omega_{A} + \frac{\gamma_{A}}{2\pi} \log\!\left(\frac{\uv - \Omega_{0}}{\Omega_{0} - \ir}\right).
    \label{eq:scalar_self_consistency_omega}
\end{align}
The parameter $\Omega_{0}$ can therefore be obtained robustly by a bisection search within the cutoff window. Once $\Omega_{0}$ is known, the second equation reduces to an algebraic equation for $\gamma_{0}$, which can be inverted analytically to obtain
\begin{align}
    \gamma_{0} = \gamma_{A} \left[ 1 + \frac{\gamma_{A}}{2\pi}\left( \frac{1}{\Omega_{0} - \ir} + \frac{1}{\uv - \Omega_{0}}\right) \right]^{-1}.
    \label{eq:bare_coupling}
\end{align}
Interestingly, recently developed time-domain simulators for this class of waveguide-QED models~\cite{bundgaard-nielsenWaveguideQEDjlEfficientFramework2025,regidorQwaveMPSEfficientOpensource2026} do not explicitly address the finite-bandwidth renormalization effects identified above. Appendix~\ref{appendix:connexion_time_domain} qualitatively illustrates that similar distortions do actually arise in time-bin simulations, although establishing their relation to the frequency-cutoff mechanism studied here would require a dedicated analysis beyond the scope of this work. We now turn to a numerical validation of Corollary~\ref{cor:renormalization_relations}.

\subsection{Beyond effective parameter renormalization}

A more general approach in scattering theory is to characterize the effective parameters $(\Omega_{\mathrm{eff}},\gamma_{\mathrm{eff}})$ through the pole $z_p$ of the atomic Green function $G_{11}(z)$ introduced in Theorem~\ref{thm:green_operator}. This approach relies on techniques from complex analysis, whose detailed treatment lies beyond the scope of the present paper but is provided in the Supplementary Material (Supplementary File). First, one can derive the self-consistent equation
\begin{align}
    \lambda_p = \sum_{n\geq 0}\alpha_n\lambda_p^n,
    \label{eq:pole_equation_waveguide}
\end{align}
where $\lambda_p=z_p-\Omega_0
=\Omega_{\mathrm{eff}}-\Omega_0-i\gamma_{\mathrm{eff}}/2$, and
\begin{align*}
\begin{dcases}
    \alpha_0
    =\frac{\gamma_0}{2\pi}
    \log\left(\frac{\Omega_0-\ir}{\uv-\Omega_0}\right)
    -i\frac{\gamma_0}{2}\\
    \alpha_{n\geq 1}
    =\frac{\gamma_0}{2n\pi}
    \left[\frac{1}{(\uv-\Omega_0)^n}
    +\frac{(-1)^{n-1}}{(\Omega_0-\ir)^n}\right]
\end{dcases}.
\end{align*}
This equation provides a systematic characterization of the effective parameters and can, in principle, be solved to arbitrary precision. In particular, its first-order truncation recovers the expressions established in Corollary~\ref{cor:renormalization_relations}.

Second, a closer examination of the analytic structure of the Green function reveals that bandwidth truncation has two distinct effects:
\begin{enumerate}[label=(\roman*)]
    \item It shifts the poles of the atomic Green function, thereby modifying the effective parameters.
    \item For very narrow bandwidths, it also distorts the functional form of the scattering amplitude, beyond what can be captured by a simple parameter renormalization.
\end{enumerate}
The cutoff-detuning assumption of Corollary~\ref{cor:renormalization_relations} identifies the regime where the former effect dominates and the latter remains negligible, as illustrated by the green region in Fig.~\ref{fig:bandwidth_reduc}. Besides offering a deeper understanding of bandwidth-truncation effects, this analysis illustrates how theoretical tools from quantum field theory can be leveraged to address practical computational challenges in quantum simulation. We now turn to a numerical validation of this calibration strategy.

\subsection{Numerical validation}

In order to numerically validate the suggested calibration procedure, we now revisit the numerical experiments of Fig.~\ref{fig:reflection_multi_cutoff_experiment_uncorrected} with the renormalization-aware parameterizations. For each frequency window, 
\begin{align*}
    [\ir, \uv] = [0, 20\pi], [0, 14\pi], [8\pi, 14\pi], [9\pi, 11\pi],
\end{align*}
we follow the inversion procedure to obtain the numerical parameters $\Omega_{0}$ and $\gamma_{0}$ supplied to the numerical model. We denote by $\mathcal{R}^{(\mathrm{corr})}$ the numerical data obtained with the corrected parameters, while $\mathcal{R}^{(\mathrm{th})}$ and $\mathcal{R}$ still denote the target prediction and the numerical data obtained with the uncorrected parameters, respectively. The additional results shown in Fig.~\ref{fig:reflection_multi_cutoff_experiment_corrected} show that the simulated atomic reflection probability has been successfully corrected to reproduce the response associated with the target parameters $\Omega_{A}$ and $\gamma_{A}$.

A small deviation remains visible in the rightmost panel of Fig.~\ref{fig:reflection_multi_cutoff_experiment_corrected}, corresponding to the narrowest window $[9\pi,11\pi]$. Near the sampled frequency-range edges, the corrected data $\mathcal{R}^{(\mathrm{corr})}$ depart from the target profile $\mathcal{R}^{(\mathrm{th})}$. This is consistent with Corollary~\ref{cor:renormalization_relations}, whose calibration formulas require frequencies sufficiently far from the cutoffs. Here, visible deviations arise when
\begin{align*}
    \min\!\left(
        |\omega_{\mathrm{init}}-\ir|,
        |\omega_{\mathrm{init}}-\uv|
    \right)
    \sim \frac{\gamma_{A}}{2}.
\end{align*}
This provides an empirical indication of when the calibration formulas may lose accuracy for the reduced-window simulations tested here: the relevant photon frequencies should remain at least of order $\gamma_A/2 = \pi/2$ away from the cutoffs. In contrast, without calibration, accurate results were obtained only with the broad reference window $[0,20\pi]$, whose cutoffs lie approximately $10\pi$ away from the probed photon frequencies. In these simulations, calibration thus reduces the required separation between the relevant photon frequencies and the cutoffs by approximately an order of magnitude.

Overall, these results support renormalization-aware calibration as a reliable route to reduced-bandwidth simulations with lower Hilbert-space dimension and computational cost. Having illustrated the corollary for arbitrary cutoffs, we now turn to symmetric windows, which provide the clearest setting and yield a compact calibration prescription in the waveguide benchmark.

\subsection{Symmetric frequency window}

The calibration relations in Eq.~\eqref{eq:renorm_relation_waveguide} simplify considerably when the frequency window is chosen symmetrically around a reference frequency. To make this explicit, we write
\begin{align*}
    \Omega_{\mathrm{ref}} = \frac{\ir + \uv}{2}, \qquad
    \Lambda = \frac{\uv-\ir}{2},
\end{align*}
so that
\begin{align*}
    [\ir,\uv]=[\Omega_{\mathrm{ref}}-\Lambda,\Omega_{\mathrm{ref}}+\Lambda].
\end{align*}
Here $\Lambda$ is the half-bandwidth, while the full numerical bandwidth is $2\Lambda$. We now take the symmetric window to be centered on the target transition frequency, $\Omega_{\mathrm{ref}}=\Omega_{A}$. In this case Eq.~\eqref{eq:scalar_self_consistency_omega} admits the simple solution
\begin{align*}
    \Omega_{0} = \Omega_{A},
\end{align*}
because the logarithmic cutoff contribution vanishes at the center of the window. This is consistent with the trend observed in subpanels (a) and (d) of Fig.~\ref{fig:reflection_multi_cutoff_experiment_uncorrected}: for the symmetric windows $[0,20\pi]$ and $[9\pi,11\pi]$, the resonance peak remained correctly centered even without corrections. The frequency calibration therefore becomes trivial: only the decay parameter has to be adjusted. Substituting $\ir=\Omega_{A}-\Lambda$, $\uv=\Omega_{A}+\Lambda$, and $\Omega_{0}=\Omega_{A}$ into Eq.~\eqref{eq:bare_coupling} gives
\begin{align*}
    \gamma_{0} = \frac{\gamma_{A}}{1 + \gamma_{A} / (\pi \Lambda)} .
\end{align*}
Thus, after the target parameters are fixed, the symmetric reduced-window calibration is controlled by a single quantity, the half-bandwidth $\Lambda$. This is the prescription stated in Algorithm~\ref{alg:symmetric_waveguide_calibration}, which also accounts for the requirement that relevant frequencies remain sufficiently far from the cutoffs. It should be used with the smallest half-bandwidth that still contains the frequencies involved in the simulated process. We now use this calibration prescription to ask whether reducing the spectral window also relaxes the time-resolution requirements of the RK4 propagation scheme.

\begin{algorithm}[t]
\caption{Symmetric-window calibration for the waveguide benchmark}
\label{alg:symmetric_waveguide_calibration}
\begin{algorithmic}[1]
\Require Target parameters $(\Omega_{A},\gamma_{A})$ and relevant photon frequencies $\{\omega_i\}_i$
\State Choose the smallest half-bandwidth $\Lambda>0$ such that 
\begin{itemize}
    \item $\{\omega_i\}_i \subset [\Omega_{A}-\Lambda,\Omega_{A}+\Lambda]$
    \item $\forall i, \quad |\omega_i - \Omega_A \pm \Lambda| \gtrsim \gamma_A / 2$
\end{itemize}
\State Set the cutoffs $\ir \gets \Omega_{A}-\Lambda$ and $\uv \gets \Omega_{A}+\Lambda$
\State Set the numerical transition frequency $\Omega_{0} \gets \Omega_{A}$
\State Set the numerical decay parameter $\displaystyle \gamma_{0} \gets \frac{\gamma_{A}}{1+\gamma_{A}/(\pi\Lambda)}$
\State Run the truncated Hamiltonian simulation with $(\ir,\uv,\Omega_{0},\gamma_{0})$
\end{algorithmic}
\end{algorithm}

\subsection{Time-step gains from bandwidth reduction}

\begin{figure}
    \centering
    \includegraphics[width=\columnwidth]{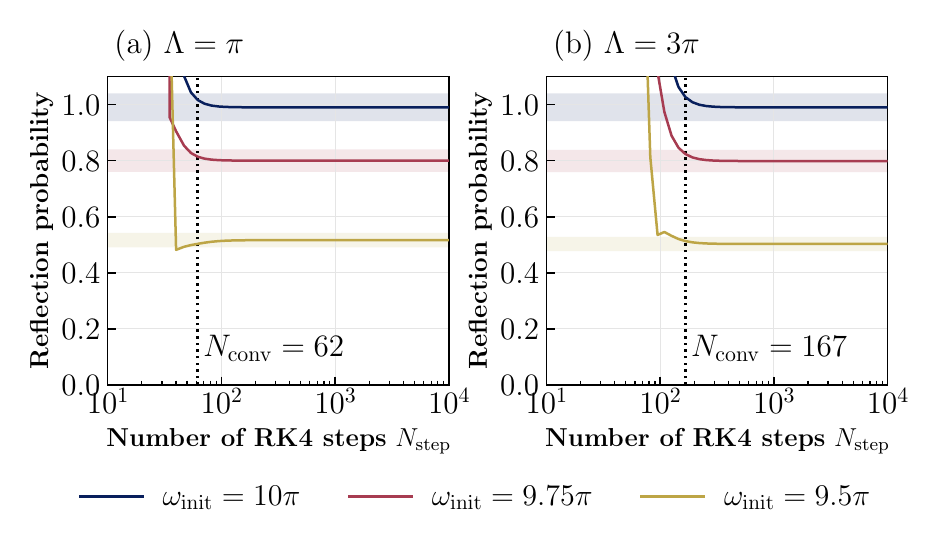}
    \caption{Reflection coefficient $\mathcal{R}$ for three incident energies as a function of the RK4 step number $N_{\mathrm{RK4}}$. The dashed lines indicate the minimum step number required for convergence.}
   \label{fig:convergence_RG_step}
\end{figure}

The motivation for this final aspect comes from a simple sampling argument. According to the Shannon sampling theorem~\cite{shannonCommunicationPresenceNoise1949}, a finite time step $\Delta t$ limits the resolvable spectral bandwidth to frequencies of order $\Delta t^{-1}$. Conversely, since in the simulation one has
\begin{align*}
    \Delta t = \frac{T}{N_{\mathrm{RK4}}},
\end{align*}
we expect a calibrated symmetric window of half-bandwidth $\Lambda$ to require a number of time steps scaling as 
\begin{align}
    N_{\mathrm{RK4}} \propto  \Lambda.
    \label{eq:nrk4_lbda_linear}
\end{align}
We test this expectation by varying $N_{\mathrm{RK4}}$ for several calibrated half-bandwidths and comparing the resulting reflection probabilities with the target profile. Figure~\ref{fig:convergence_RG_step} shows representative results for $\Lambda=\pi$ and $\Lambda=3\pi$, corresponding respectively to the previously considered windows $[9\pi,11\pi]$ and $[7\pi,13\pi]$, for the three incident frequencies used in the reflection benchmark. For each bandwidth, we define the converged step number $N_{\mathrm{RK4}}^{\mathrm{conv}}$ as the smallest value for which all three probed photon frequencies remain within a $5\%$ error band around their target values. This tolerance is used here as a simple indicative accuracy criterion, providing a practical reference point for comparing bandwidths. The reduced window reaches this criterion with fewer RK4 steps.

\begin{figure}
    \centering
    \includegraphics[width=0.9\linewidth]{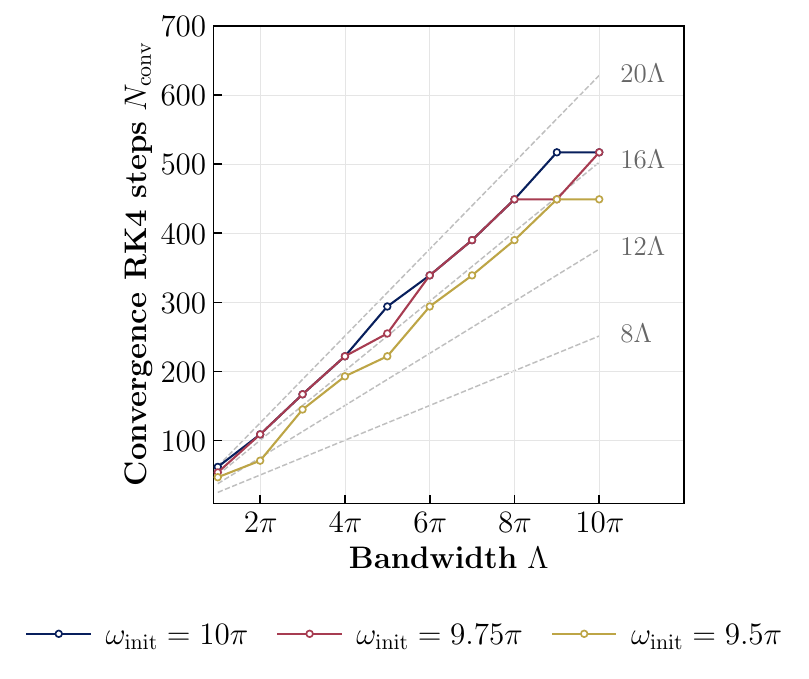}
    \caption{Minimum number of RK4 time steps required for convergence as a function of the half-bandwidth $\Lambda$.}
   \label{fig:Nconv_vs_bandwidth}
\end{figure}

Repeating the extraction over several half-bandwidths reveals a linear trend shown in Fig.~\ref{fig:Nconv_vs_bandwidth}, consistent with Eq.~\eqref{eq:nrk4_lbda_linear}. In the single-excitation waveguide benchmark, the number of retained photonic modes scales as $N_{\mathrm{modes}}=\mathcal{O}(\Lambda/\Delta p)$ at fixed momentum spacing. Since the vectorized RK4 cost is linear in both $N_{\mathrm{RK4}}$ and $N_{\mathrm{modes}}$, the calibrated symmetric-window simulations therefore scale as
\begin{align*}
    \mathcal{O}\!\left(N_{\mathrm{RK4}}^{\mathrm{conv}}\,\frac{\Lambda}{\Delta p}\right)
    =
    \mathcal{O}\!\left(\frac{\Lambda^2}{\Delta p}\right),
\end{align*}
where the last equality uses the empirical linear dependence. Thus, reducing the half-bandwidth from $\Lambda=10\pi$ to $\Lambda=\pi$ lowers both the number of retained modes and the required RK4 step number by about one order of magnitude in this one-dimensional benchmark. With the indicative $5\%$ accuracy criterion used here, this corresponds to a reduction of the propagation cost by about two orders of magnitude while preserving the calibrated scattering response. This estimate is specific to the present waveguide benchmark and should not be considered as a complete performance analysis. Nevertheless, obtaining such a gain in this simple setting suggests that renormalization-aware bandwidth reduction is a promising route for more demanding photon-matter simulations. A systematic investigation of these gains beyond the waveguide benchmark is left for future work.

\section{Extending calibration to a two-dimensional atomic array}
\label{sec:two_dimensional_calibration}

We now extend the single-photon calibration strategy to a two-dimensional model whose scattering response is not analytically tractable in general. We consider the atomic array introduced earlier, with
\begin{align*}
    N_A = 10, \quad d = 2, \quad \omega_{\bm{p}} = \left\lVert \bm{p} \right\rVert_2, \quad g_{0,j}(\bm{p}) = i d_{0,j} e^{-i\bm{p}\cdot \bm{x}_j}.
\end{align*}
We define an operational target response from a large-window reference simulation, without assuming that it reproduces the exact continuum response. Our objective is then to assess whether a reduced-bandwidth model, calibrated using Theorem~\ref{thm:green_operator}, can consistently reproduce this prescribed response.

\subsection{Atomic-array scattering setup}

The simulated optical device is a finite slab made of $N_A=10$ identical TLSs. The atoms have the same target transition frequency $\Omega_{A}$ and the same target coupling strength $d_{A,j}$, and their positions $\bm{x}_j$ are arranged along a straight line whose axis is tilted by $45^\circ$ with respect to the horizontal propagation direction (see Fig.~\ref{fig:model_calibration_summary}). This simple geometry is meant to retain the microscopic character of the atomic-array models of Refs.~\cite{havukainenQuantumSimulationsOptical1999,obaFastSimulationMultiphoton2024}. The atomic coordinates used here are provided in the geometry file \texttt{geometry\_id5.csv} in the accompanying code repository~\cite{pironGitHub}.

The incoming photon is prepared to the left of the array, with initial position centered near the line $y=0$. Its central momentum points toward the atomic slab,
\begin{align*}
    \bm{p}_{\mathrm{init}} = (\omega_{\mathrm{init}},0),
\end{align*}
so that the wave packet initially propagates along the positive $x$ direction. In this experiment, we do not attempt to probe an entire spectral profile, as we did for the reflection response in the waveguide experiment. Indeed, because no analytical baseline is available, the accuracy of the resulting spectral profile cannot be assessed directly. Instead, for selected incident frequencies, we test whether the simulated response remains consistent across different numerical bandwidths. Specifically, we consider $\omega_{\mathrm{init}} \in \{8\pi,9\pi,10\pi\}$.

We quantify the effective response by the evolution of the transmitted wave packet. Following the microscopic optical-element simulations of Refs.~\cite{havukainenQuantumSimulationsOptical1999,obaFastSimulationMultiphoton2024}, we define the mode fidelity as the overlap between the initial and current wave-packet amplitudes in the interaction picture. Its continuous-time definition and its discrete counterpart evaluated at the simulation time steps $t_i=i\Delta t$ are given by
\begin{align}
    \begin{dcases}
        \mathcal{F}_{\mathrm{mode}}(t) &= |\langle \psi(t) | \psi(0) \rangle|^2 \\
        \mathcal{F}_{\mathrm{mode},i} &= |\langle \psi_i | \psi_0 \rangle|^2
    \end{dcases}.
\end{align}
Therefore, any measured deformation can be attributed to the photon-matter interaction. In the absence of atoms, one expects the final mode fidelity to satisfy $\mathcal{F}_{\mathrm{mode}, N_{\mathrm{step}}}=1$. By contrast, if the slab prevents the wave packet from continuing along its initial propagation path, as in the case of a perfectly reflecting slab, then $\mathcal{F}_{\mathrm{mode}, N_{\mathrm{step}}}=0$. We now detail the initialization of the system and illustrate the resulting dynamics through a representative scattering example.

\subsection{Two-dimensional wave-packet initialization}

\begin{figure*}
    \centering
    \includegraphics[width=\linewidth]{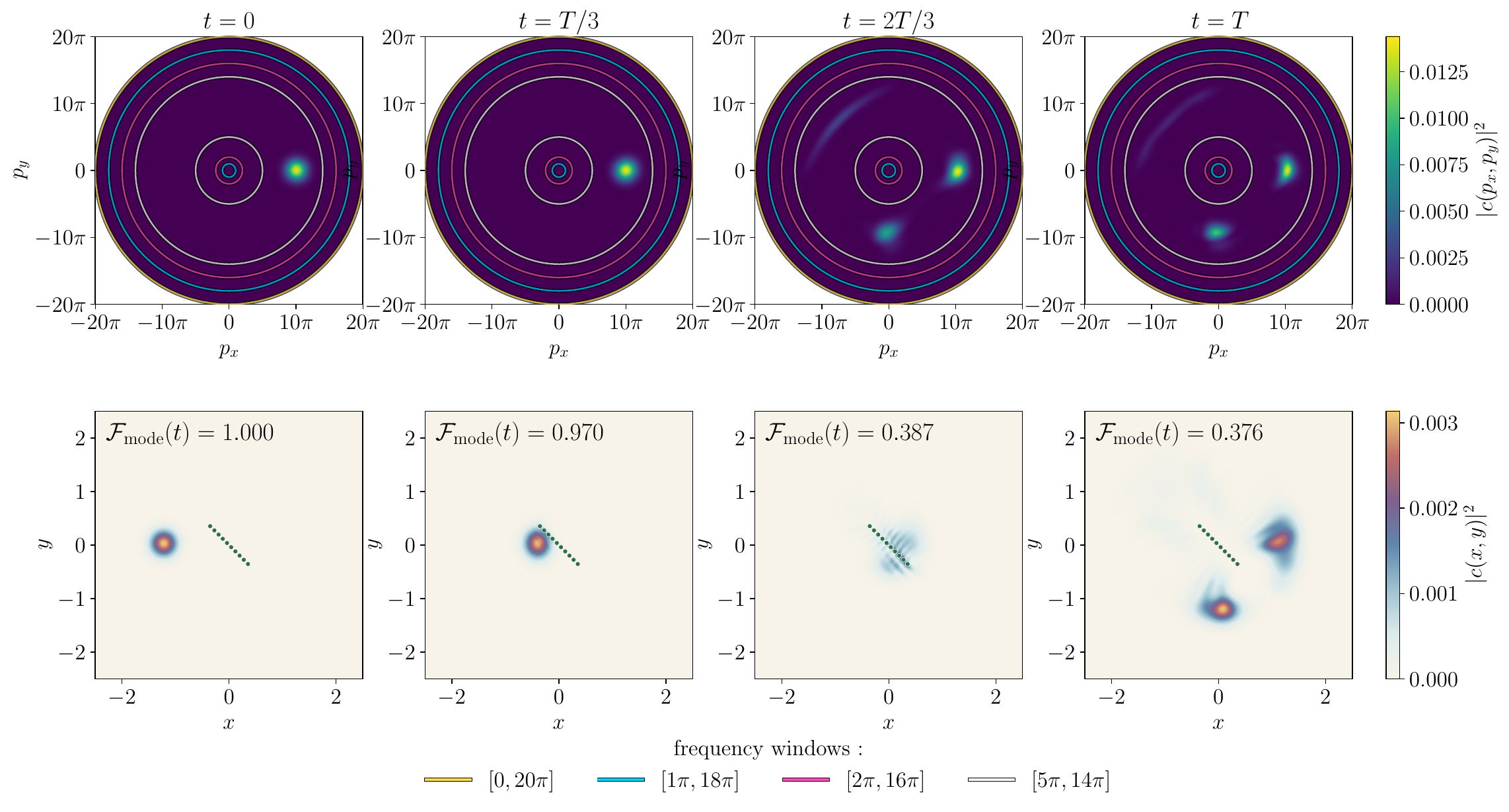}
    \caption{Single-photon probability during interaction with the two-dimensional atomic array, shown in momentum (top) and position (bottom) space. Position-space probabilities are obtained via the discrete Fourier transform (Appendix~\ref{appendix:discrete_position_transform}), with $\mathcal{F}_{\mathrm{mode}}(t)$ values reported on the corresponding plots. The reference window $[0,20\pi]$ is used, with the occupied momentum support remaining well within its bounds.}
   \label{fig:density_snapshots_2d}
\end{figure*}

We now specify the single-photon simulation used to probe this two-dimensional setup. We first fix target values $(\Omega_{A,j}, d_{A,j})$ for the common transition frequencies and coupling strengths of the TLSs. The initial state contains one photon on the two-dimensional momentum grid, while all atoms are initially in their ground state. We use a Gaussian wave packet with central momentum $\bm{p}_{\mathrm{init}}$, momentum widths $(\sigma_{p,x},\sigma_{p,y})$, and initial real-space center $\bm{x}_{\mathrm{init}}$:
\begin{align*}
    \ket{\psi_0}
    =
    \sum_{\bm{n}\in\mathcal{I}_{\ir,\uv}}
    c_{\bm{n},0}\ket{\bm{p}_{\bm{n}}},
\end{align*}
with coefficients
\begin{align*}
    \begin{split}
    c_{\bm{n},0}
    \propto
    \exp & \Bigg[ -\frac{(p_{n_x}-p_{\mathrm{init},x})^2}{4\sigma_{p,x}^2} -\frac{(p_{n_y}-p_{\mathrm{init},y})^2}{4\sigma_{p,y}^2} \\
    & \hspace*{6em} -i\bm{p}_{\bm{n}}\cdot\bm{x}_{\mathrm{init}}\Bigg].
    \end{split}
\end{align*}
As in the waveguide benchmark, the proportionality constant is fixed numerically so that the state is normalized within the retained finite frequency window. The main simulation parameters used below are summarized in Table~\ref{tab:simulation_parameters_2d}. In particular, we assume uniform atomic parameters,
\begin{align*}
    \begin{dcases}
        \Omega_{A,j} &= \Omega_A = 10\pi\\
        d_{A,j} &= d_A = 1
    \end{dcases},
    \qquad j\in\{1,\dots,N_A\}.
\end{align*}

\begin{table}[h]
\centering
\begin{tabular}{|l l|}
\hline
\multicolumn{2}{|c|}{\textbf{Target TLS parameters}} \\
\hline
Atomic frequency $\Omega_{A,j}$ &  $10\pi$ \\
Target coupling $d_{A,j}$ & $1$ \\
\hline
\multicolumn{2}{|c|}{\textbf{Photon parameters}} \\
\hline
Initial energy $\omega_{\mathrm{init}}$ & $\{10\pi, 9\pi, 8\pi\}$ \\
Central momentum $\bm{p}_{\mathrm{init}}$ & $(\omega_{\mathrm{init}},0)$ \\
Momentum spread $(\sigma_{p,x}, \sigma_{p,y})$ & $(\pi, \pi)$ \\
Initial position $\bm{x}_{\mathrm{init}}$ & $(-1.25,0)$ \\
\hline
\multicolumn{2}{|c|}{\textbf{Numerical simulation parameters}} \\
\hline
Number of atoms $N_A$ &  $10$ \\
Lengths $(L_1, L_2)$ & $(5,5)$ \\
Simulation time $T$ & $2.5$ \\
RK4 time step $\Delta t$ & $0.01$ \\
Numerical frequencies of the TLSs $\Omega_{0,j}$ & $10\pi$ \\
Numerical couplings $d_{0,j}$ & $1$ \\
\hline

\end{tabular}
\vspace*{1em}
\caption{Parameters used in the atomic array simulations}
\label{tab:simulation_parameters_2d}
\end{table}

\subsection{Bandwidth of reference}

For the reference simulation, inspired by the waveguide benchmark, we use the frequency window
\begin{align}
    [\ir^{\mathrm{ref}}, \uv^{\mathrm{ref}}] = [0, 20\pi].
\end{align}
It is centered around the common target atomic frequency $\Omega_A=10\pi$, and provided satisfactory accuracy even before calibrating the model in the waveguide benchmark. We therefore expect the occupied momentum modes to remain well within the region defined by these reference cutoffs throughout the evolution. To verify this, Fig.~\ref{fig:density_snapshots_2d} shows the corresponding dynamics at selected times $t_i \in \{0, T/3, 2T/3, T\}$. The upper panels display the discrete momentum-mode probabilities $|c_{\bm{n},i}|^2$, while the lower panels show the corresponding position-space probabilities obtained through the discrete Fourier transform described in Appendix~\ref{appendix:discrete_position_transform}.

The broad window $[0,20\pi]$ defines a large disk in the two-dimensional momentum plane. Throughout the scattering process, the populated modes remain concentrated away from both its outer boundary and the low-frequency region near the origin, suggesting that narrower windows may still capture the momentum-space support relevant to the dynamics. We therefore consider three representative reduced windows: $[\pi,18\pi]$, $[2\pi,16\pi]$, and $[5\pi,14\pi]$. Taking the simulation over $[0,20\pi]$ as the reference, we investigate whether calibration based on Theorem~\ref{thm:green_operator} allows these reduced-window models to reproduce the same mode fidelity.

\subsection{Spectral matching-based calibration}

In the waveguide benchmark, the analytical expression for the scattering amplitude $\mathcal{T}_{\ir,\uv}$ allowed us to derive an explicit calibration condition by matching the reduced-bandwidth response to a prescribed analytical target. In the present two-dimensional model, no such closed-form expression is generally available. We therefore formulate the calibration directly in terms of the projected Green functions, requiring the reduced-bandwidth spectral response to match that of the reference simulation. This leads to the following corollary.

\begin{corollary}[Atomic array calibration]
\label{cor:two_dimensional_spectral_matching}

Consider the two-dimensional model above with fixed atomic positions $\{\bm{x}_j\}_{j=1}^{N_A}$. Let a reference response be defined by 
\begin{align*}
    \begin{dcases}
        [\ir^{\mathrm{ref}},\uv^{\mathrm{ref}}] & \text{broad reference frequency window}\\
        \Omega_{A,j} & \text{target atomic frequencies}\\
        d_{A,j} & \text{target couplings}
    \end{dcases}.
\end{align*}
For a reduced window $[\ir,\uv]\subseteq[\ir^{\mathrm{ref}},\uv^{\mathrm{ref}}]$, with numerical parameters $(\Omega_{0,j},d_{0,j})$, define the reduced and target spectral profiles
\begin{align}
    \begin{split}
    \mathcal{F}_{jk}^{(0)}(\omega)
    &=
    d_{0,j}^*
    \left[\mathsf{G}_{\mathrm{eff}}^{(\ir,\uv)}(\omega + i0^+)\right]_{jk}
    d_{0,k},\\
    \mathcal{F}_{jk}^{(A)}(\omega)
    &=
    d_{A,j}^*
    \left[\mathsf{G}_{\mathrm{ref}}^{(\ir^{\mathrm{ref}},\uv^{\mathrm{ref}})}(\omega + i0^+)\right]_{jk}
    d_{A,k},
    \end{split}
    \label{eq:two_dimensional_target_profile}
\end{align}
where $\mathsf{G}_{\mathrm{eff}}^{(\ir,\uv)}$ and $\mathsf{G}_{\mathrm{ref}}^{(\ir^{\mathrm{ref}},\uv^{\mathrm{ref}})}$ respectively denote the atomic Green matrices for the reduced and reference frequency windows. Explicitly,
\begin{align*}
    \begin{split}
    \mathsf{G}_{\mathrm{eff}}^{(\ir,\uv)}(\omega + i0^+)
    &=
    \Big[
    \omega\mathsf{I}_{N_A} - \mathsf{\Omega}_{0} \\
    &\qquad
    - \mathsf{\Sigma}_{\mathrm{eff}}^{(\ir,\uv)}(\omega + i0^+)
    \Big]^{-1}, \\
    \mathsf{G}_{\mathrm{ref}}^{(\ir^{\mathrm{ref}},\uv^{\mathrm{ref}})}(\omega + i0^+)
    &=
    \Big[
    \omega\mathsf{I}_{N_A} - \mathsf{\Omega}_{A} \\
    &\qquad
    - \mathsf{\Sigma}_{\mathrm{ref}}^{(\ir^{\mathrm{ref}},\uv^{\mathrm{ref}})}(\omega + i0^+)
    \Big]^{-1}.
    \end{split}
\end{align*}
Here $\mathsf{\Omega}_{s} = \operatorname{Diag}(\bm{\Omega}_{s})$, with $\bm{\Omega}_{s}=(\Omega_{s,1},\ldots,\Omega_{s,N_A})^{\mathsf{T}}$ for $s\in\{0,A\}$, and the reference self-energy is evaluated using the target couplings $d_{A,j}$. A sufficient condition for reproducing the reference scattering amplitude is then
\begin{align}
    \begin{split}
    \mathcal{F}_{jk}^{(0)}(\omega)
    &=
    \mathcal{F}_{jk}^{(A)}(\omega),\\
    &\qquad
    \forall \omega\in[\ir,\uv],\ \forall j,k.
    \end{split}
    \label{eq:two_dimensional_spectral_profile}
\end{align}

\begin{proof}
See Appendix~\ref{appendix:proof_corollary_2}.
\end{proof}
\end{corollary}

The question is then how to achieve this spectral agreement in practice. The waveguide benchmark revealed two distinct effects of bandwidth reduction: a shift in the effective scattering parameters and a distortion of the spectral shape of the scattering amplitude, as further detailed analytically in the Supplementary Material. For the two-dimensional array, we similarly expect moderate bandwidth reductions to primarily shift the effective parameters while approximately preserving the spectral shape. Rather than analytically characterizing the regime in which such distortions remain negligible, we investigate empirically whether the spectral matching condition of Corollary~\ref{cor:two_dimensional_spectral_matching} can be approximately satisfied by adjusting the numerical parameters $(\bm{\Omega}_0,\bm{d}_0)$. Specifically, taking $[\ir^{\mathrm{ref}},\uv^{\mathrm{ref}}]=[0,20\pi]$, we test this calibration strategy for three reduced windows,
\begin{align}
    [\ir,\uv]\in
    \left\{
        [\pi,18\pi],
        [2\pi,16\pi],
        [5\pi,14\pi]
    \right\}.
\end{align}
Our objective is thus to assess the practical effectiveness of parameter adjustments for these representative windows, without claiming a general characterization of the calibration strategy's range of validity.

\begin{algorithm}[t]
\caption{Spectral-matching calibration for the two-dimensional array}
\label{alg:spectral_matching_calibration}
\begin{algorithmic}[1]
\Require Target parameters $(\Omega_{A,j},d_{A,j})$, reference cutoffs $[\ir^{\mathrm{ref}},\uv^{\mathrm{ref}}]$, reduced cutoffs $[\ir,\uv]$, and calibration frequencies $\mathcal{D}$
\State Compute the reference Green matrix $\mathsf{G}_{\mathrm{ref}}^{(\ir^{\mathrm{ref}},\uv^{\mathrm{ref}})}(\omega_{\ell}+i0^+)$ for each $\omega_{\ell}\in\mathcal{D}$
\State Build the target spectral profiles $\mathcal{F}_{\ell,jk}\gets \mathcal{F}_{jk}^{(A)}(\omega_{\ell})$
\State Choose initial numerical parameters $(\Omega_{0,j},d_{0,j})$ for the reduced-window model
\State Minimize $\mathcal{L}^{\mathrm{cal}}_{\ir,\uv}$ over $(\Omega_{0,j},d_{0,j})$
\State Run the reduced-window simulation with the optimized numerical parameters
\end{algorithmic}
\end{algorithm}

\begin{figure}
    \centering
    \includegraphics[width=0.9\linewidth]{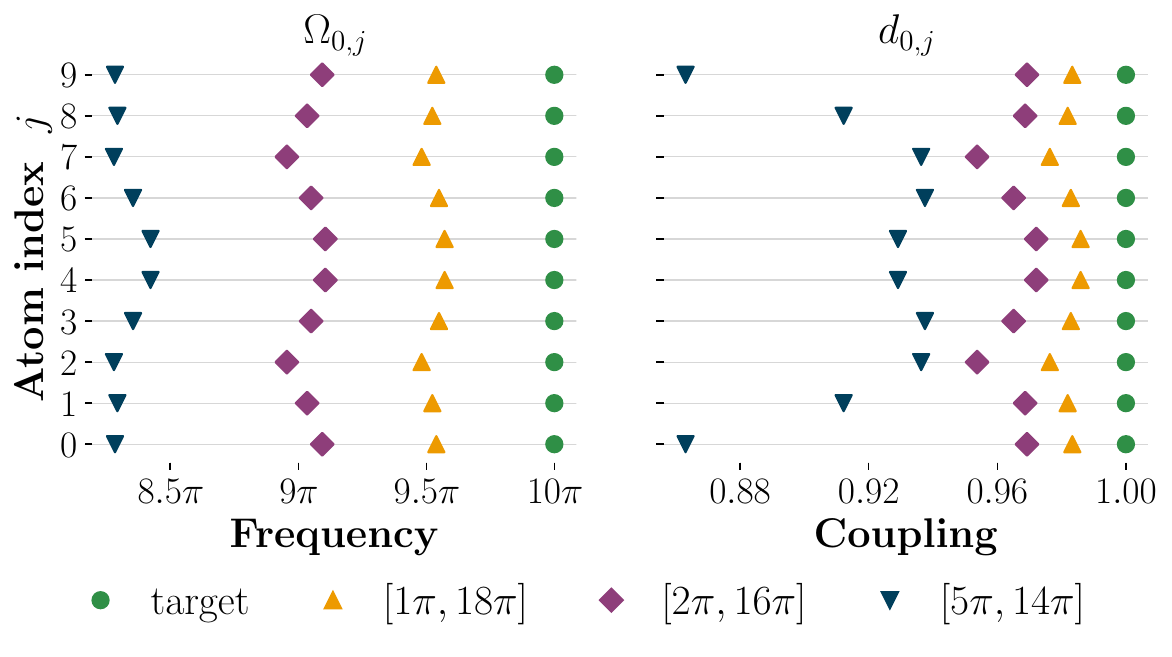}
    \caption{Numerical transition frequencies and couplings obtained by spectral matching for the reduced frequency windows.}
   \label{fig:bare_parameters_freqwindows_2d}
\end{figure}

\begin{figure*}
    \centering
    \includegraphics[width=0.95\linewidth]{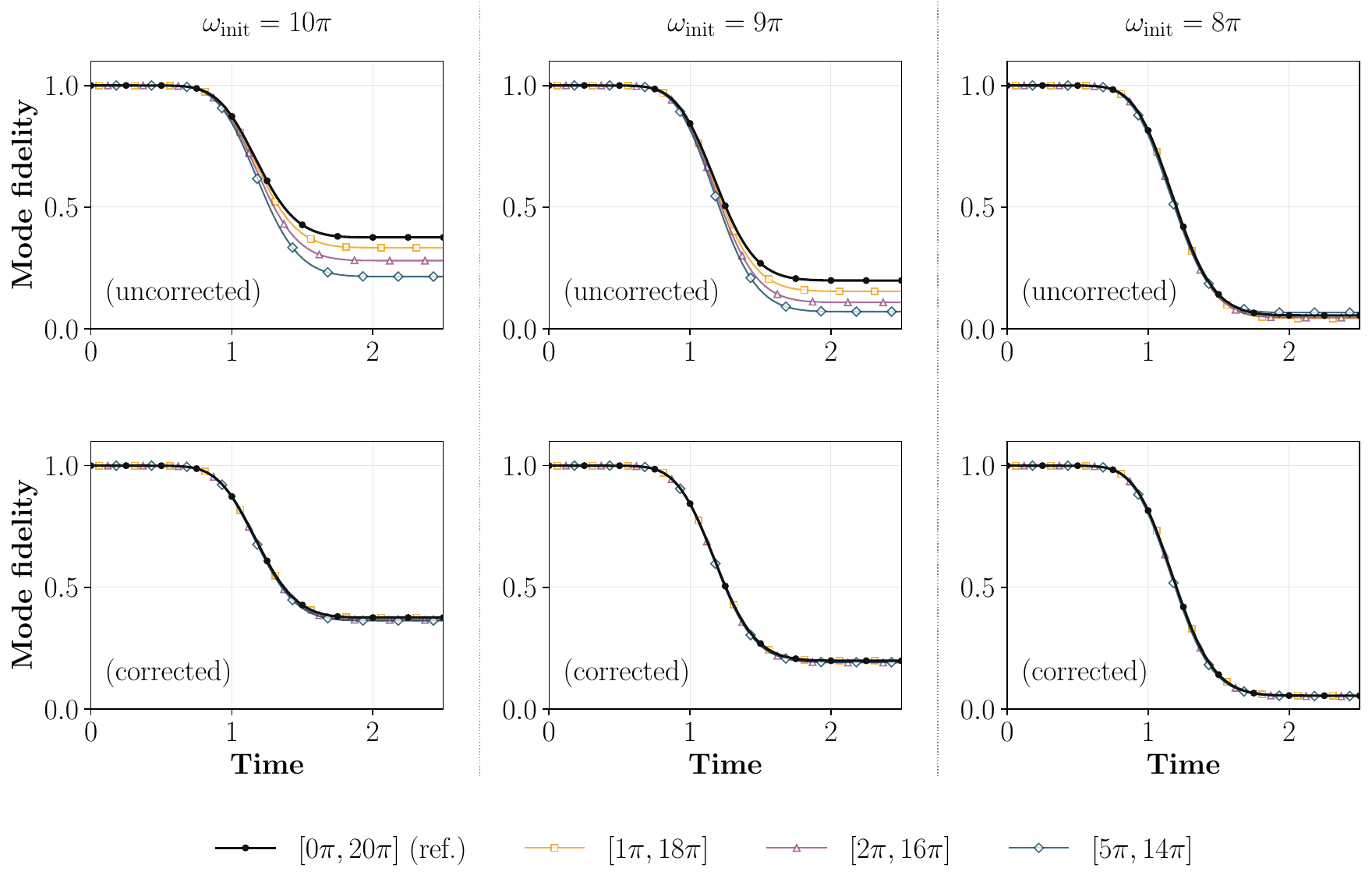}
    \caption{Time-dependent mode fidelity for uncalibrated and calibrated reduced-window simulations.}
   \label{fig:timedep_modefidelity_vs_freqwindow_2d}
\end{figure*}

To this end, we formulate spectral matching as a least-squares optimization problem over the numerical parameters of the reduced-window model. We begin by discretizing the broad reference window into a regular finite calibration grid,
\begin{align*}
    \mathcal{D} = \left\{\omega_{\ell}\right\}_{\ell=1,\dots,N_{\mathrm{cal}}} \subset [\ir^{\mathrm{ref}},\uv^{\mathrm{ref}}].
\end{align*}
For a chosen reduced window $[\ir,\uv]$, only the calibration frequencies that remain inside the reduced window are used.
At each $\omega_{\ell}\in\mathcal{D} \cap [\ir, \uv]$, we evaluate both the reduced-window spectral profile and the reference target profile,
\begin{align*}
\begin{dcases}
    \mathcal{F}_{\ell,jk}^{(0)}
    =
    \mathcal{F}_{jk}^{(0)}(\omega_{\ell})\\
    \mathcal{F}_{\ell,jk}^{(A)}
    =
    \mathcal{F}_{jk}^{(A)}(\omega_{\ell})
\end{dcases}.
\end{align*}
The residual for a trial reduced-window parameterization is then
\begin{align*}
    \Delta_{\ell,jk}(\{\Omega_{0,j}\}, \{d_{0,j}\})
    =
    \mathcal{F}_{\ell,jk}^{(0)}
    -
    \mathcal{F}_{\ell,jk}^{(A)}, 
\end{align*}
and the calibration is obtained by solving the unweighted least-squares problem
\begin{align*}
    \min_{\{\Omega_{0,j}\},\{d_{0,j}\}}
    \mathcal{L}^{\mathrm{cal}}_{\ir,\uv}(\{\Omega_{0,j}\},\{d_{0,j}\}),
\end{align*}
with
\begin{align*}
    \mathcal{L}^{\mathrm{cal}}_{\ir,\uv}(\{\Omega_{0,j}\},\{d_{0,j}\}) =
    \sum_{\ell}
    \sum_{j,k=1}^{N_A} \left| \Delta_{\ell,jk} \right|^2.
\end{align*}
\texttt{scipy.optimize.least\_squares} is used in the numerical implementation below to solve this least-squares problem. The resulting workflow is summarized in Algorithm~\ref{alg:spectral_matching_calibration}.

\subsection{Consistency across reduced windows}
We now apply the spectral-matching procedure of Algorithm~\ref{alg:spectral_matching_calibration} to the three reduced windows identified above,
\begin{align*}
    [\pi, 18\pi], \qquad [2\pi, 16\pi], \qquad [5\pi, 14\pi].
\end{align*}
The reference interval $[0,20\pi]$ is sampled on a regular grid of $N_{\mathrm{cal}}=300$ frequencies, which defines the calibration grid $\mathcal{D}$. For each reduced-window fit, the optimization is initialized at the uncalibrated target parameters,
\begin{align*}
    (\bm{\Omega}_{0}, \bm{d}_{0})_{\mathrm{init}}
    =
    (\bm{\Omega}_{A}, \bm{d}_{A}).
\end{align*}
We use \texttt{scipy.optimize.least\_squares} with bounded search intervals for the numerical transition frequencies and couplings,
\begin{align*}
    \Omega_{0,j} \in [\ir, \uv],
    \qquad
    d_{0,j} \in [-\infty, +\infty],
\end{align*}
and with the stopping tolerances
\begin{align*}
    \begin{dcases}
        \texttt{ftol} = 10^{-8} & \text{(cost-function tolerance)}\\
        \texttt{xtol} = 10^{-8} & \text{(parameter-step tolerance)}\\
        \texttt{gtol} = 10^{-8} & \text{(gradient tolerance)}
    \end{dcases}.
\end{align*}
The resulting numerical parameters are reported in Fig.~\ref{fig:bare_parameters_freqwindows_2d}. This calibration step does not require propagating the full photon-matter state in the broad reference window: it only requires evaluating the finite-window Green kernels entering Eq.~\eqref{eq:two_dimensional_target_profile}. Once the optimized parameters have been found for a chosen frequency window, they are kept fixed throughout the subsequent wave-packet propagation. The least-squares fit can therefore be performed offline, before running the actual dynamical simulation, so that the reduced-window propagation retains the lower Hilbert-space cost.

The effect of this calibration is shown in Fig.~\ref{fig:timedep_modefidelity_vs_freqwindow_2d}. Without calibration, reducing the frequency window changes the time-dependent mode fidelity in a window-dependent way. The discrepancy is moderate for the lowest incident frequency, $\omega_{\mathrm{init}}=8\pi$, but is clearly visible for $\omega_{\mathrm{init}}=9\pi$ and $\omega_{\mathrm{init}}=10\pi$, where the uncalibrated reduced-window evolutions no longer reproduce the broad-window reference. After spectral matching, the different reduced windows collapse back onto the reference curve over the whole interaction time. This demonstrates that Green-kernel calibration closely reproduces the reference mode-fidelity dynamics for the tested incident wave packets and reduced windows, indicating that the bandwidth-induced distortions of the scattering dynamics have been effectively compensated for.

This two-dimensional example provides a proof of concept for applying Green-kernel calibration beyond analytically solvable waveguide models. A detailed optimization study of the associated computational gain, including the interplay between bandwidth reduction and time-step requirements in higher dimensions, is left for future work.

\section{Conclusion}

In this work, we studied how frequency truncation can be handled rigorously in numerical simulations of quantum photon-matter models, with the broader motivation of supporting the design of quantum-network components. We summarize the main findings below and discuss possible next steps.

\subsection{Summary of contributions}

We have made four main contributions.
\begin{itemize}
    \item Starting from a simple waveguide benchmark, we identified that reduced-bandwidth photon-matter simulations are not straightforward to set up. Even when the frequency window contains the relevant energy scales of the problem, reducing the bandwidth can modify physical observables in an unexpected way, through effective shifts that are directly visible in the scattering dynamics.
    \item We used elements of scattering theory to formulate a precise framework for this effect. This led to a single-photon consistency condition, expressed in terms of the Green-matrix response, which can be used as a calibration condition for finite-bandwidth simulations.
    \item We presented two applications of this calibration idea: the waveguide benchmark model and a two-dimensional atomic-array model. In both cases, the calibrated reduced-bandwidth simulations restore consistency at the single-photon level.
    \item Using the simple symmetric-window prescription in the waveguide benchmark, we showed that reducing the bandwidth by a factor of ten decreases both the number of retained modes and the number of RK4 steps required for accurate propagation. This results in an overall reduction in computational cost by approximately a factor of $100$, even in the single-photon sector of this minimal model.
\end{itemize}

\subsection{Future work}

Several directions remain open. In particular, it would be interesting to extend the spectral-matching calibration to broader classes of photon-matter models that are analytically challenging to study, including, but not limited to,
\begin{itemize}
    \item Multiple emitters coupled to the same waveguide~\cite{sheremetWaveguideQuantumElectrodynamics2022,pletyukhovScatteringMasslessParticles2012}, for which virtual-photon exchange between emitters may introduce additional cutoff-dependent effects
    \item Emitters with frequency-dependent couplings and structured spectral densities, as encountered in non-Markovian and circuit-QED settings~\cite{malekakhlaghCutofffreeCircuitQuantum2017,domokosQuantumDescriptionLight2002,gonzalez-gutierrezNonMarkovianDynamicsGiant2025}.
\end{itemize}
Finally, the calibration strategy developed here should be tested beyond the single-photon sector. Although calibrating the one-photon response may not be sufficient to fully reproduce nonlinear multiphoton dynamics, correcting the single-photon contribution embedded in the multiphoton response may already lead to significant improvements. Quantifying the extent of these improvements would help determine how far reduced-bandwidth, renormalization-aware simulations can be pushed as a practical tool for modeling quantum-optical devices.

\section*{Acknowledgment}
R.P. acknowledges the hospitality of the National Institute of Informatics during the research stay that laid the groundwork for this project. OpenAI's ChatGPT, using the GPT-5.5 model, was used for language editing and stylistic refinement throughout the manuscript, to generate portions of the simulation code used for the two-dimensional results presented in Section~\ref{sec:two_dimensional_calibration}, and to assist with the editing and preparation of all figures. All AI-assisted outputs were reviewed, verified, and, where necessary, modified by the authors.

\section*{Code availability}
The numerical code and atomic-geometry data are available in the public GitHub repository~\cite{pironGitHub} of the project.

\appendix

\bibliographystyle{IEEEtran}
\bibliography{library}


\appendices

\makeatletter
\renewcommand{\p@subsection}{\thesection-}
\makeatother

\section{Vectorized single-photon propagation scheme}
\label{appendix:detailed_rk_scheme}

Here, we detail the numerical implementation of the RK4 scheme introduced in Eqs.~\eqref{eq:rk4_propagation} and~\eqref{eq:rk4_increment_states}. Under the rotating-wave approximation (RWA), the total excitation number is conserved. Since the simulations presented in this work involve a single photon propagating in vacuum, we restrict the numerical propagation to the single-excitation sector. We consider a generic spatial dimension $d$ and an arbitrary number $N_A$ of two-level atoms, covering both the waveguide and two-dimensional configurations investigated in this work. Starting from a single-photon state with all atoms in their ground state, the system remains in the single-excitation sector, where its state at time $t_i$ reads
\begin{align}
    \ket{\psi_i}
    =
    \sum_{\bm{n}\in\mathcal{I}_{\ir,\uv}}
    c_{\bm{n},i}\ket{\bm{p}_{\bm{n}}}
    +
    \sum_{j=1}^{N_A} b_{j,i}\ket{e_j}.
    \label{eq:psi_i_coordinates}
\end{align}
Here, $\ket{\bm{p}_{\bm{n}}}$ denotes a single photon in momentum mode $\bm{p}_{\bm{n}}$, with all atoms in their ground state, while $\ket{e_j}$ denotes an excitation of atom $j$ in the absence of photons. Collecting the photonic and atomic amplitudes into vectors $\bm{c}_i$ and $\bm{b}_i$, respectively, the state is represented numerically by
\begin{align*}
    \bm{\psi}_i
    =
    \begin{pmatrix}
        \bm{c}_i\\
        \bm{b}_i
    \end{pmatrix},
    \quad
    \bm{c}_i
    =
    \begin{pmatrix}
        c_{\bm{n}_1,i}\\
        \vdots\\
        c_{\bm{n}_{N_{\mathrm{modes}}},i}
    \end{pmatrix},
    \quad
    \bm{b}_i
    =
    \begin{pmatrix}
        b_{1,i}\\
        \vdots\\
        b_{N_A,i}
    \end{pmatrix},
\end{align*}
where $\bm{n}_1,\dots,\bm{n}_{N_{\mathrm{modes}}}$ enumerate the $N_{\mathrm{modes}}$ retained momentum modes in $\mathcal{I}_{\ir,\uv}$. Applying the interaction Hamiltonian $V(t)$ defined in Eq.~\eqref{eq:interacting_hamilt} to the state decomposition in Eq.~\eqref{eq:psi_i_coordinates}, we represent $V(t)$ by its matrix $\mathsf{V}(t)$ in the ordered basis $(\ket{\bm{p}_{\bm{n}_1}},\ldots,\ket{\bm{p}_{\bm{n}_{N_{\mathrm{modes}}}}},\ket{e_1},\ldots,\ket{e_{N_A}})$ and obtain
\begin{align}
    \mathsf{V}(t)\bm{\psi}_i
    =
    \begin{pmatrix}
        \mathsf{G}(t)\bm{b}_i\\
        \mathsf{G}^\dagger(t)\bm{c}_i
    \end{pmatrix},
    \label{eq:appendix-vectorized-interaction}
\end{align}
where $\mathsf{G}(t)\in\mathbb{C}^{N_{\mathrm{modes}}\times N_A}$ is the time-dependent coupling matrix with elements
\begin{align*}
    G_{\bm{n},j}(t)
    =
    g_{\bm{n},j}^*
    e^{i(\omega_{\bm{p}_{\bm{n}}}-\Omega_j)t}.
\end{align*}
For completeness, the RK4 propagation over a time step $\Delta t$ is
\begin{align*}
    \bm{\psi}_{i+1}
    &=
    \bm{\psi}_i+\delta\bm{\psi}_i,\\
    \delta\bm{\psi}_i
    &=
    \frac{
    \delta\bm{\psi}_{i1}
    +2\delta\bm{\psi}_{i2}
    +2\delta\bm{\psi}_{i3}
    +\delta\bm{\psi}_{i4}}{6},
\end{align*}
with
\begin{align}
    \begin{dcases}
    \delta\bm{\psi}_{i1}
    &=-i\Delta t\,\mathsf{V}(t_i)\bm{\psi}_i\nonumber\\
    \delta\bm{\psi}_{i2}
    &=-i\Delta t\,\mathsf{V}\left(t_i+\frac{\Delta t}{2}\right)
    \left(\bm{\psi}_i+\frac{\delta\bm{\psi}_{i1}}{2}\right)\nonumber\\
    \delta\bm{\psi}_{i3}
    &=-i\Delta t\,\mathsf{V}\left(t_i+\frac{\Delta t}{2}\right)
    \left(\bm{\psi}_i+\frac{\delta\bm{\psi}_{i2}}{2}\right)\nonumber\\
    \delta\bm{\psi}_{i4}
    &=-i\Delta t\,\mathsf{V}(t_i+\Delta t)
    \left(\bm{\psi}_i+\delta\bm{\psi}_{i3}\right)
    \end{dcases}.
\end{align}
At each intermediate stage, the action of $\mathsf{V}(t)$ is evaluated using Eq.~\eqref{eq:appendix-vectorized-interaction}, involving only matrix-vector products between the coupling matrix and the photonic or atomic amplitudes.

\section{Technical details for the waveguide benchmark}

This appendix covers the following technical details:
\begin{itemize}
    \item Derivation of the waveguide self-energy.
    \item Proof of Corollary~\ref{cor:renormalization_relations}.
    \item Discussion of the connection between the frequency-domain approach developed here and time-domain simulation methods.
\end{itemize}

\subsection{Waveguide self-energy}
\label{appendix:waveguide_self_energy}
We begin with the general expression for the self-energy in Eq.~\eqref{eq:general_self_energy}:
\begin{align*}
    \mathsf{\Sigma}(z) = \int d\omega' \frac{\mathsf{J}(\omega')}{z-\omega'}.
\end{align*}
Since the waveguide contains a single atom, both the spectral density and the self-energy reduce to scalars. In one dimension with $\omega_p=|p|$, the two momenta at fixed energy $\omega>0$ are $p=\pm\omega$. The spectral density is therefore
\begin{align*}
    \begin{split}
        J(\omega) &= \int \frac{dp}{2\pi}\, g_0(p)g_0^*(p) \delta(\omega-\omega_p) \\
        &\hspace*{4cm}= \frac{|g_0(\omega)|^2+|g_0(-\omega)|^2}{2\pi}.
    \end{split}
\end{align*}
In the flat-coupling testbed, $g_{0}(p)=i\sqrt{\gamma_{0}/2}$, and hence
\begin{align*}
    J(\omega)=\frac{\gamma_{0}}{2\pi}.
\end{align*}
To obtain the effective amplitude at an arbitrary frequency $\omega$, we evaluate the self-energy at $z = \omega + i0^+$. The Sokhotski-Plemelj formula reads
\begin{align*}
    \frac{1}{\omega-\omega'+i0^+} = \mathcal{P}\frac{1}{\omega-\omega'} - i\pi\delta(\omega-\omega'),
\end{align*}
where $\mathcal{P}$ denotes the Cauchy principal value. For the waveguide model, this gives, for $\omega \in [\ir, \uv]$,
\begin{align*}
    \begin{split}
    &\Sigma_{\mathrm{eff}}^{(\ir,\uv)}(\omega+i0^+) \\
    &= \frac{\gamma_0}{2\pi} \mathrm{P.V.}\left[\int_{\ir}^{\uv} \frac{d\omega'}{\omega-\omega' + i0^+}\right] - i \frac{\gamma_0}{2} \\
    &= \frac{\gamma_0}{2\pi} \log \left( \frac{\omega - \ir}{\uv - \omega} \right) - i\frac{\gamma_0}{2}.
    \end{split}
\end{align*}

\begin{figure*}
    \centering
    \includegraphics[width=\linewidth]{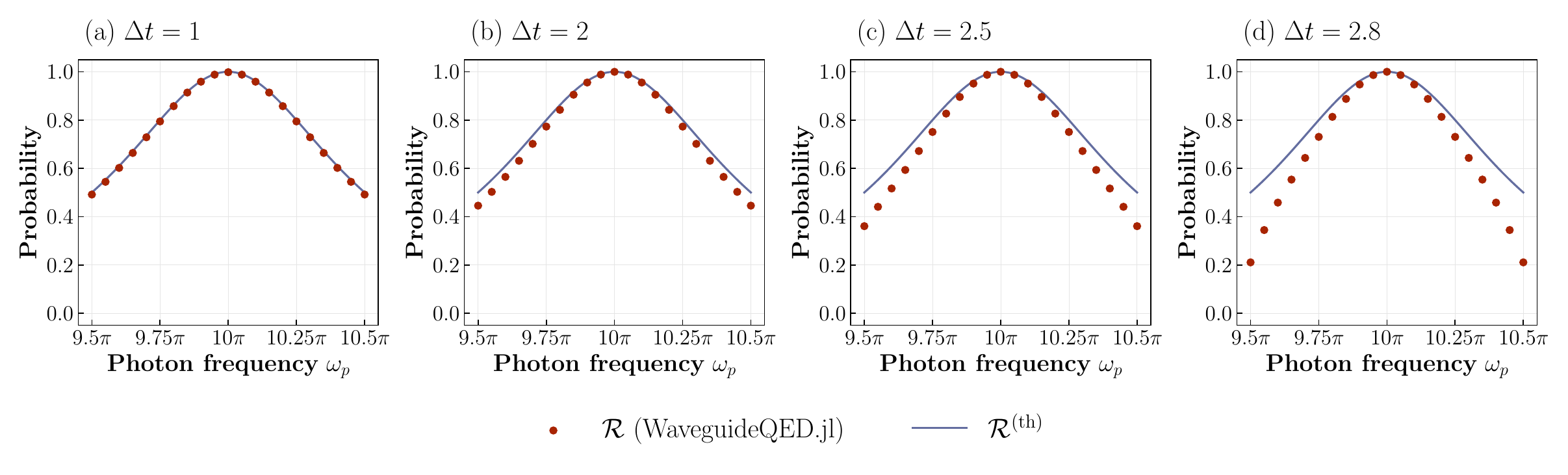}
    \caption{Reflection profiles computed with \texttt{WaveguideQED.jl} for increasing time steps, illustrating effective-parameter shifts reminiscent of finite-bandwidth effects in momentum-space simulations.}
   \label{fig:reflection_profile_jl}
\end{figure*}

\subsection{Proof of Corollary 1}
\label{appendix:proof_crly1}

Consider an incoming photon propagating to the right with momentum $p>0$ and energy $\omega=\omega_p=p$. The reflected component has momentum $-p$, so the amplitude introduced in Theorem~\ref{thm:green_operator} can be identified with the reflection amplitude:
\begin{align*}
    \mathcal{T}_{\ir,\uv}(-p,p;\omega) = r^{\mathrm{eff}}_{\omega}, \qquad \omega=p.
\end{align*}
Applying Theorem~\ref{thm:green_operator} to the single-TLS benchmark gives
\begin{align*}
    r^{\mathrm{eff}}_{\omega} = -\frac{i \gamma_{0}/2}{\omega-\Omega_{0}-\Sigma_{\mathrm{eff}}^{(\ir,\uv)}(\omega+i0^+)}.
\end{align*}
Using the effective self-energy derived above, we recover exactly the expression stated in the main text. We then expand the self-energy around $\omega=\Omega_{0}$:
\begin{align*}
    \Sigma_{\mathrm{eff}}^{(\ir,\uv)}(\omega+i0^+) = \sum_{n\geq 0} \alpha_n(\omega-\Omega_{0})^n ,
\end{align*}
where the Taylor coefficients $\alpha_n$ are defined by
\begin{align*}
    \alpha_n = \frac{1}{n!}\left.\frac{\partial^n \Sigma_{\mathrm{eff}}^{(\ir,\uv)}(\omega+i0^+)}{\partial \omega^n}\right|_{\omega=\Omega_{0}}.
\end{align*}
The $n=0$ term follows by evaluating the self-energy at the numerical TLS frequency. For $n \geq 1$, repeated differentiation followed by evaluation at $\omega=\Omega_{0}$ gives the exact coefficients:
\begin{align*}
    \begin{dcases}
    \alpha_0 &= \frac{\gamma_{0}}{2\pi}\log\!\left( \frac{\Omega_{0}-\ir}{\uv-\Omega_{0}} \right) - i \frac{\gamma_{0}}{2} \\
    \alpha_{ n\geq 1} &= \frac{\gamma_{0}}{2n \pi}\left[\frac{(-1)^{n-1}}{(\Omega_{0}-\ir)^n}+\frac{1}{(\uv-\Omega_{0})^n}\right]
    \end{dcases}.
\end{align*}
These coefficients coincide with those in the self-consistent pole equation~\eqref{eq:pole_equation_waveguide} in the main text. The triangle inequality gives the bound
\begin{align*}
    |\alpha_n| \leq \frac{\gamma_{0}}{2n\pi}\left(\frac{1}{|\Omega_{0}-\ir|^n}+\frac{1}{|\uv-\Omega_{0}|^n}\right), \quad n\geq 1.
\end{align*}
Hence, provided the cutoffs lie sufficiently far from the numerical TLS frequency $\Omega_{0}$, the terms with $n\geq 2$ can be neglected, yielding
\begin{align*}
    \begin{split}
        & \omega-\Omega_{0}-\Sigma_{\mathrm{eff}}^{(\ir,\uv)}(\omega+i0^+) \\
        & \hspace*{9em}\simeq -\alpha_0 + (1-\alpha_1)(\omega-\Omega_{0}).
    \end{split}
\end{align*}
Substituting this expression into the effective amplitude yields
\begin{align*}
    r_{\omega}^{\mathrm{eff}} = -\frac{i \gamma_0/(2(1-\alpha_1))}{\omega - \Omega_0 - \alpha_0/(1-\alpha_1)}.
\end{align*}
Because $\alpha_1$ is real and $\Im(\alpha_0) = -\gamma_0/2$, we obtain
\begin{align*}
    r_{\omega}^{\mathrm{eff}} = -\frac{i \gamma_0/(2(1-\alpha_1))}{\omega - \Omega_0 - \Re(\alpha_0)/(1-\alpha_1) +i \gamma_0/(2(1-\alpha_1))}.
\end{align*}
This has the functional form stated in Corollary~\ref{cor:renormalization_relations}, with
\begin{subequations}
\begin{align*}
    \Omega_{\mathrm{eff}} &= \Omega_{0} + \frac{\Re(\alpha_0)}{1-\alpha_1}, \\
    \gamma_{\mathrm{eff}} &= \frac{\gamma_{0}}{1-\alpha_1}.
\end{align*}
\end{subequations}
Substituting the explicit expressions for $\alpha_0$ and $\alpha_1$ recovers the effective parameters in Corollary~\ref{cor:renormalization_relations}, completing the proof. The same result follows from solving the self-consistent pole equation~\eqref{eq:pole_equation_waveguide} truncated at first order.

\subsection{Connection with time-domain simulators}
\label{appendix:connexion_time_domain}

The waveguide-QED model has also been extensively studied analytically using the so-called input-output formalism~\cite{fanInputOutputFormalismFewPhoton2010}, which relies on a time-domain formulation of the photon-matter interaction. This formalism has provided the theoretical basis for efficient time-domain solvers~\cite{heuckPhotonPhotonInteractionsDynamically2020} for this waveguide-QED model, such as \texttt{WaveguideQED.jl}~\cite{bundgaard-nielsenWaveguideQEDjlEfficientFramework2025} and \texttt{QWaveMPS}~\cite{regidorQwaveMPSEfficientOpensource2026}. It is therefore natural to ask whether effects similar to those identified in this work also persist in time-domain simulators. We argue that they do.

A finite time step $\Delta t$ introduces an effective frequency scale $|\uv-\ir|\sim \Delta t^{-1}$ through the Nyquist cutoff associated with the Shannon sampling theorem~\cite{shannonCommunicationPresenceNoise1949}. It may therefore induce cutoff-dependent shifts in the effective scattering parameters. This suggests that some apparent inaccuracies arising when the time step is increased could potentially be mitigated by a renormalization-aware calibration procedure.

This qualitative behavior is illustrated in Fig.~\ref{fig:reflection_profile_jl}, where we compute the same reflection profile using the time-domain package \texttt{WaveguideQED.jl}~\cite{bundgaard-nielsenWaveguideQEDjlEfficientFramework2025} while progressively increasing the time step. The resulting profiles display symptoms reminiscent of the cutoff-induced distortions studied above. Interestingly, the observed shift tends to narrow the reflection profile, whereas the finite-frequency-window effects considered in this work broaden it. This difference emphasizes that the correspondence is only qualitative: time bins are not literally equivalent to the frequency windows used here, and a dedicated analysis would be required to establish the precise relation between time-domain discretization and effective renormalization.

\section{Technical details for the two-dimensional model}

This appendix covers the following technical details:
\begin{itemize}
    \item Construction of the position-space plots in Fig.~\ref{fig:density_snapshots_2d}.
    \item Derivation of the atomic array self-energy.
    \item Proof of Corollary~\ref{cor:two_dimensional_spectral_matching}.
\end{itemize}

\subsection{Discrete Fourier transform for position-space visualization}
\label{appendix:discrete_position_transform}

The position-space panels in Fig.~\ref{fig:density_snapshots_2d} display discrete probability distributions associated with the finite set of momentum amplitudes stored in the simulation. We describe the transform below to specify the normalization convention. At time $t_i$, the photon-sector state in the retained momentum window is represented by coefficients $c_{\bm{n}}(t_i) = c_{\bm{n},i}$, with $\bm{n}=(n_x,n_y)\in\mathcal{I}_{\ir,\uv}$. Since the RK4 evolution is performed in the interaction picture, the Schrödinger-picture coefficients used for position-space visualization are
\begin{align*}
    \widetilde{c}_{\bm{n}}(t_i)
    =
    c_{\bm{n}}(t_i)e^{-i\omega_{\bm{n}}t_i}.
\end{align*}
The retained set $\mathcal{I}_{\ir,\uv}$ is selected by a radial frequency window and is therefore not generally rectangular. For the discrete Fourier transform, we embed it in the smallest centered rectangular grid
\begin{align*}
    R
    =
    \{-N_x^{\max},\ldots,N_x^{\max}\}
    \times
    \{-N_y^{\max},\ldots,N_y^{\max}\},
\end{align*}
where $N_x^{\max}=\max_{\bm{n}\in\mathcal{I}_{\ir,\uv}}|n_x|$ and similarly for $N_y^{\max}$. We then define
\begin{align*}
    M_x=2N_x^{\max}+1,\qquad
    M_y=2N_y^{\max}+1,
\end{align*}
and sample the periodic simulation box on the dual position grid
\begin{align*}
    x_a=-\frac{L_1}{2}+\frac{aL_1}{M_x},
    \qquad
    y_b=-\frac{L_2}{2}+\frac{bL_2}{M_y},
\end{align*}
with $a=0,\ldots,M_x-1$ and $b=0,\ldots,M_y-1$. The discrete position coefficients used in Fig.~\ref{fig:density_snapshots_2d} are then
\begin{align}
    c_{a,b}(t_i)
    =
    \frac{1}{\sqrt{M_xM_y}}
    \sum_{\bm{n}\in\mathcal{I}_{\ir,\uv}}
    \widetilde{c}_{\bm{n}}(t_i)
    e^{i\bm{p}_{\bm{n}}\cdot (x_a,y_b)} ,
    \label{eq:discrete_position_transform}
\end{align}
and the plotted quantity is $P_{a,b}(t)=|c_{a,b}(t)|^2$. This restricted transform is an isometry, and therefore
\begin{align*}
    \sum_{a,b} |c_{a,b}(t)|^2
    =
    \sum_{\bm{n}\in\mathcal{I}_{\ir,\uv}}
    |c_{\bm{n}}(t)|^2 .
\end{align*}
The position-space panels therefore represent normalized discrete probability maps on the finite Fourier grid.

\subsection{Atomic array self-energy}
We now derive the finite-bandwidth self-energy for the two-dimensional model. The code uses this self-energy to compute the projected resolvent, from which the spectral functions used for calibration by spectral matching are obtained. For $\omega>0$, the spectral-density matrix is
\begin{align*}
    \begin{split}
        J_{jk}(\omega) &= \int \frac{d^2\bm{p}}{(2\pi)^2} g_j(\bm{p}) g_k^*(\bm{p}) \delta(\omega - \omega_{\bm{p}}) \\
        & =
        \frac{\omega d_{0,j} d_{0,k}^*}{4\pi^2}
        \int_0^{2\pi} d\theta\,
        e^{-i\omega \bm{x}_{jk} \cdot \hat{\bm{n}}_{\theta}} .
    \end{split}
\end{align*}
Here, the unit vector $\hat{\bm{n}}_{\theta}=(\cos\theta,\sin\theta)$ specifies the momentum-space direction in polar coordinates, while $\bm{x}_{jk}=\bm{x}_j-\bm{x}_k$ denotes the relative position vector between atoms $j$ and $k$. The angular integral can be expressed in terms of a Bessel function. Writing $\bm{x}_{jk} = (r_{jk} \cos \phi, r_{jk} \sin \phi)$ and shifting the angle $\theta \rightarrow \theta - \phi$ gives
\begin{align*}
    \int_0^{2\pi} d\theta\, e^{-i\omega \bm{x}_{jk} \cdot \hat{\bm{n}}_{\theta}} 
    =
    \int_{0}^{2\pi} d\vartheta \, e^{-i \omega r_{jk} \cos \vartheta} 
    = 
    2 \pi \mathrm{J}_0(\omega r_{jk}),
\end{align*}
where $\mathrm{J}_0$ is the Bessel function of the first kind~\cite{micheliIntegralRepresentationBessels2017}. The spectral-density matrix is therefore
\begin{align*}
    J_{jk}(\omega) = d_{0,j}d_{0,k}^* \frac{\omega}{2\pi} \mathrm{J}_0(\omega r_{jk}).
\end{align*}
As in the waveguide model, applying the Sokhotski-Plemelj formula gives the self-energy at $z = \omega + i0^+$:
\begin{align*}
    \begin{split}
    & \Sigma_{\mathrm{eff},jk}^{(\ir,\uv)}(\omega+i0^+) \\
    & \hspace*{2em}= \frac{d_{0,j} d_{0,k}^*}{2\pi}\,
    \mathcal{I}_{jk}^{(\ir,\uv)}(\omega) 
    - i\,\frac{\omega}{2}d_{0,j} d_{0,k}^* \mathrm{J}_0(\omega r_{jk}),
    \end{split}
\end{align*}
where the band-limited principal-value integral is defined as
\begin{align*}
    \mathcal{I}_{jk}^{(\ir,\uv)}(\omega)
    &=
    \mathrm{P.V.}\Bigg[
    \int_{\ir}^{\uv} d\omega'\,
    \frac{\omega'\mathrm{J}_0(\omega' r_{jk})}{\omega-\omega'}
    \Bigg].
\end{align*}
Here, P.V. denotes the Cauchy principal value.

\subsection{Proof of Corollary 2}
\label{appendix:proof_corollary_2}

Energy conservation restricts the incoming and outgoing momenta to the same energy, so the effective amplitude $\mathcal{T}_{\ir, \uv}$ should be evaluated between momenta of the form 
\begin{align*}
    \begin{dcases}
    \bm{p}_{\phi} &=\omega\hat{\bm{n}}_{\phi}\\
    \bm{q}_{\theta} &=\omega\hat{\bm{n}}_{\theta}
    \end{dcases},
    \quad\text{where } \hat{\bm{n}}_{\alpha}=(\cos\alpha,\sin\alpha).
\end{align*}
The effective amplitude is then
\begin{align*}
    \begin{split}
    &\mathcal{T}_{\ir,\uv}(\theta,\phi,\omega;\{\Omega_{0,j}\},\{d_{0,j}\}) \\
    & \hspace*{5em} = -i\,\bm{g}^{\dagger}(\bm{q}_{\theta}) \mathsf{G}_{\mathrm{eff}}^{(\ir,\uv)}(\omega + i0^+) \bm{g}(\bm{p}_{\phi}),
    \end{split}
\end{align*}
and the calibration condition can be written formally as
\begin{align*}
    \mathcal{T}_{\ir,\uv}(\theta,\phi,\omega;\{\Omega_{0,j}\},\{d_{0,j}\})
    \simeq
    f_{\mathrm{target}}(\theta,\phi,\omega).
\end{align*}
In this two-dimensional model, we do not assume that the target profile is available in closed analytical form. Instead, we construct it by evaluating $\mathcal{T}$ for a broad reference window with the target parameters:
\begin{align*}
    f_{\mathrm{target}}(\theta, \phi, \omega) = \mathcal{T}_{\ir^{\mathrm{ref}},\uv^{\mathrm{ref}}}\left(\theta,\phi,\omega;\{\Omega_{A,j}\}, \{d_{A,j}\}\right).
\end{align*}
Let us now develop the mathematical expression of the amplitude $\mathcal{T}$. At fixed $\omega$, the coupling vector is
\begin{align*}
    \bm{g}(\bm{p}_{\phi})
    =
    \begin{pmatrix}
        i d_{0,1} e^{-i\omega \hat{\bm{n}}_{\phi}\cdot \bm{x}_1}\\
        \vdots\\
        i d_{0,N_A} e^{-i\omega \hat{\bm{n}}_{\phi}\cdot \bm{x}_{N_A}}
    \end{pmatrix}.
\end{align*}
Hence, for a given set of numerical parameters $(\bm{\Omega}_0, \bm{d}_0)$ and a frequency window $[\ir, \uv]$, the corresponding effective amplitude is given by
\begin{align*}
    \begin{split}
    &\mathcal{T}_{\ir,\uv}(\theta,\phi,\omega;\{\Omega_{0,j}\},\{d_{0,j}\}) \\
    & \hspace*{8em} =  -i \sum_{j,k} \mu_{jk}(\theta,\phi,\omega) \mathcal{F}^{(0)}_{jk}(\omega),
    \end{split}
\end{align*}
where
\begin{align*}
    \begin{split}
        \mu_{jk}(\theta,\phi,\omega) &= e^{i\omega \hat{\bm{n}}_{\theta}\cdot\bm{x}_j} e^{-i\omega \hat{\bm{n}}_{\phi}\cdot\bm{x}_k}, \\
        \mathcal{F}^{(0)}_{jk}(\omega) &= d_{0,j}^* \left[\mathsf{G}_{\mathrm{eff}}^{(\ir,\uv)}(\omega + i0^+)\right]_{jk} d_{0,k}.
    \end{split}
\end{align*}
Since the calibrated and reference configurations have the same atomic positions, the angular factors $\mu_{jk}$ are identical in both effective and reference amplitudes. The matching condition therefore reads
\begin{align*}
    \sum_{j,k} \mu_{jk}(\theta,\phi,\omega) \left[ \mathcal{F}^{(0)}_{jk}(\omega) - \mathcal{F}^{(A)}_{jk}(\omega) \right] = 0,
\end{align*}
which is satisfied for all $\theta,\phi$ whenever Eq.~\eqref{eq:two_dimensional_spectral_profile} holds, completing the proof.

\clearpage
\onecolumn
\begingroup
\restoreSection
\setcounter{section}{0}
\setcounter{subsection}{0}
\setcounter{equation}{0}
\setcounter{figure}{0}
\setcounter{table}{0}
\renewcommand{\thesection}{S\Roman{section}}
\renewcommand{\thesectiondis}{\thesection.}
\renewcommand{\thesubsection}{\Alph{subsection}}
\renewcommand{\thesubsectiondis}{\thesubsection.}
\renewcommand{\theequation}{S\arabic{equation}}
\renewcommand{\thefigure}{S\arabic{figure}}
\renewcommand{\thetable}{S\arabic{table}}
\renewcommand{\theHsection}{supp.\arabic{section}}
\renewcommand{\theHsubsection}{\theHsection.\arabic{subsection}}
\renewcommand{\theHequation}{supp.\arabic{equation}}
\renewcommand{\theHfigure}{supp.\arabic{figure}}
\renewcommand{\theHtable}{supp.\arabic{table}}
\linespread{1.06}\selectfont
\setlength{\parskip}{3pt plus 1pt minus 1pt}
\begin{center}
{\Large\bfseries Supplementary material\par}
\medskip
{\large Bandwidth-Cost Reduction for Photon-Matter Simulation at the Single-Photon Level\par}
\end{center}

This supplementary note presents analytical developments explored during the preparation of the manuscript. Owing to the particularly simple structure of the waveguide model considered in the main text, these developments are not required to derive the effective parameters used there. Nevertheless, the framework presented here is more general and may provide a useful route for analyzing bandwidth-reduction effects in more complex light-matter models. While the individual ingredients are well established in the literature, their presentation in this context also illustrates how concepts and analytical tools developed in quantum field theory and renormalization theory can be used to address practical numerical-simulation problems in quantum engineering.

Here, we focus on the waveguide configuration with a single atom.

\section*{Conventions}

\begin{itemize}
    \item For a complex-valued function, we use the boundary-value notation
        \begin{align}
            \begin{dcases}
                f(x + i0^+) &= \lim_{\epsilon \rightarrow 0^+} f(x + i \epsilon)\\
                f(x + i0^-) &= \lim_{\epsilon \rightarrow 0^-} f(x + i \epsilon)
            \end{dcases}
        \end{align}
    \item We use the principal branch of the complex logarithm,
\begin{align}
        \log z = \log |z| + i \operatorname{Arg} z, \quad \operatorname{Arg} z \in (-\pi, \pi]
\end{align}
    with the branch cut along the negative real axis:
    \begin{align}
        \begin{dcases}
            \log(x + i0^+) &= \log(-x) + i\pi \\
            \log(x - i0^+) &= \log(-x) - i\pi \\
            \log(x + i0^+) - \log(x - i0^+) &= 2i\pi
        \end{dcases}, x<0
    \end{align}
\end{itemize}

\section{Pole analysis of the atomic Green function}

Theorem 1 of the main text shows that, in the single-photon sector, the scattering response emulated by the simulation is fully determined by the nontrivial part of the scattering matrix,
\begin{align}
    \mathcal{T}_{\ir,\uv}(\bm{q},\bm{p};\omega)
    =
    -i\,\bm{g}^{\dagger}(\bm{q})
    \bm{G}_{\mathrm{eff}}(\omega+i0^+)
    \bm{g}(\bm{p}),
\end{align}
where $\bm{G}_{\mathrm{eff}}(z)$ is the band-limited atomic Green's function. In this section, we examine why its resonance poles are related to the effective parameters of the waveguide model. Let us temporarily omit the subscript ``eff'' for clarity. For a single atom, the Green's function reduces to a scalar and, as derived in the main text and in Ref.~\cite{schneiderGreensfunctionFormalismWaveguide2016}, reads
\begin{align}
    G_{11}(z) = \frac{1}{z-\Omega_0 -\Sigma(z)}, \qquad \Sigma(z) = \frac{\gamma_0}{2\pi} \log\left(\frac{z - \ir}{z - \uv}\right), \qquad \operatorname{Im}z>0,
    \label{supp:eq:self_energy_def}
\end{align}
where $\Sigma(z)$ is the self-energy. The resonance pole $z_p$ is obtained by solving
\begin{align}
    D(z_p) = 0, \qquad D(z) = z - \Omega_0 - \Sigma(z).
\end{align}
Since Eq.~\eqref{supp:eq:self_energy_def} is stated for $\operatorname{Im}z>0$, locating a decaying resonance requires an analytic continuation of $\Sigma$ into the lower half-plane. We construct this continuation explicitly below when deriving the self-consistent pole equation. In the meantime, $G_{11}$ and $D$ evaluated near $z_p$, including the derivative $D'(z_p)$, are understood in terms of this continuation.

\subsection{Analytical structure of the atomic Green function}

In order to understand the physical significance of the resonance pole, let us first write the Green in an appropriate form. We perform a Taylor expansion of the denominator around the pole,
\begin{align}
    D(z) = \sum_{n \geq 1} (z-z_p)^n \frac{1}{n!} D^{(n)}(z_p).
\end{align}
Then, assuming that the pole is simple, i.e. $D'(z_p) \ne 0$, the Green function takes the form
\begin{align}
    G_{11}(z) = \frac{1}{D'(z_p)(z-z_p)} \Bigg[ 1 + \sum_{n\geq 1} (z-z_p)^n \frac{D^{(n+1)}(z_p)}{(n+1)!\,D'(z_p)} \Bigg]^{-1} = \frac{1}{D'(z_p)(z-z_p)} + G_{11,\mathrm{reg}}(z),
    \label{supp:eq:pole_decomposition}
\end{align}
where we used the Taylor series $(1+x)^{-1} = 1 - x + x^2 - \dots$ near $x=0$. In particular, $G_{11,\mathrm{reg}}(z)$ is regular at $z=z_p$. Notice that this decomposition can also be obtained using a Laurent expansion of $G_{11}(z)$ around the pole. Having identified this analytic structure, let us show how it enters the atomic dynamics.

\subsection{Physical interpretation of the pole}

To interpret the pole, we now consider the atomic dynamics in the single-excitation sector. Assume that the atom is initially excited, and the field is in vacuum. The system then evolves according to the Schrödinger equation,
\begin{align}
    \begin{dcases}
        \ket{\psi(0)} &= \ket{0,e} \\
        \partial_t \ket{\psi(t)} &= -i H \ket{\psi(t)}
    \end{dcases}
\end{align}
At time $t$, the state in the Schrödinger picture is given by 
\begin{align}
   \ket{\psi(t)} = e^{-iHt} \ket{0,e} = \int \frac{dp}{2\pi} c(p,t)\ket{p,g} + b(t)\ket{0,e},
\end{align}
where we used excitation number conservation to expand the state in the single excitation basis, and where the integral is restricted to the retained photonic modes. In particular, the atomic amplitude is
\begin{align}
    b(t) = \langle 0, e | \psi(t) \rangle =  \bra{0,e}e^{-iHt}\ket{0,e}
\end{align}
Its Laplace transform $\widetilde b(s) = \mathcal{L}\Big[t \mapsto b(t)\Big](s)$ is defined, for $\operatorname{Re}s>0$, by
\begin{align}
    \widetilde b(s)= \int_0^\infty dt\,e^{-st}b(t)
    = \Bigl\langle 0,e \Bigm| \int_{0}^{\infty}\! dt\; e^{\,-(s + iH)t} \Bigm| 0,e \Bigr\rangle =  \Bigl\langle 0,e \Bigm| \frac{1}{s + iH}  \Bigm| 0,e \Bigr\rangle
    =iG_{11}(is).
    \label{supp:eq:atomic_laplace_transform}
\end{align}
Thus the same Green function governs scattering and atomic decay~\cite{domokosQuantumDescriptionLight2002,scullyQuantumOptics1997}. Accordingly, the original amplitude is recovered by the inverse Laplace transform, for $t>0$, 
\begin{align}
    b(t)= \mathcal{L}^{-1}\Big[ s \mapsto iG_{11}(is) \Big] (t) = \mathcal{L}^{-1}\Bigg[ \frac{1}{D'(z_p)(s+iz_p)} \Bigg] + \mathcal{L}^{-1}\Big[ i G_{11, \mathrm{reg}}(is) \Big],
    \label{supp:eq:atomic_inverse_laplace}
\end{align}
where we used the linearity of the inverse Laplace transform. Using the formula
\begin{align}
    \mathcal{L}^{-1}\Bigg[ \frac{1}{(s+iz_p)} \Bigg](t) = e^{-iz_p t},
\end{align}
one ends up with the exact decomposition
\begin{align}
    b(t)=\frac{e^{-iz_pt}}{D'(z_p)}+b_{\mathrm{reg}}(t), \qquad b_{\mathrm{reg}}(t)
    =\mathcal{L}^{-1}\Big[ i G_{11, \mathrm{reg}}(is) \Big].
\end{align}
Here $b_{\mathrm{reg}}$ is the remainder after extracting the selected resonance pole. Writing $z_p=\Omega_{\mathrm{eff}}-i\gamma_{\mathrm{eff}}/2$ makes the physical interpretation explicit:
\begin{align}
    b(t)=\frac{e^{-i\Omega_{\mathrm{eff}}t}e^{-\gamma_{\mathrm{eff}}t/2}}{D'(z_p)}
       +b_{\mathrm{reg}}(t).
\end{align}
When $\gamma_{\mathrm{eff}}>0$, the first term oscillates at $\Omega_{\mathrm{eff}}$, and its squared modulus decays as $e^{-\gamma_{\mathrm{eff}}t}$. These parameters therefore characterize the effective frequency and the effective decay rate entering the atomic dynamics. The full amplitude retains $b_{\mathrm{reg}}(t)$ and need not follow a purely exponential decay law.

\subsection{Self-consistent equation for the pole}

We now seek a resonance pole $z_p = \Omega_{\mathrm{eff}} - i\gamma_{\mathrm{eff}}/2$ corresponding to a decaying contribution, with its effective frequency lying between the cutoffs. We thus require
\begin{align}
    \ir < \Omega_{\mathrm{eff}} < \uv, \qquad \gamma_{\mathrm{eff}} > 0.
\end{align}
Such a pole lies in the lower half-plane, $\operatorname{Im}z<0$, whereas Eq.~\eqref{supp:eq:self_energy_def} defines the retarded self-energy for $\operatorname{Im}z>0$. We therefore need to extend that definition to $\operatorname{Im}z<0$.

Let $\Sigma_{\mathrm{I}}(z)$ denote the expression in Eq.~\eqref{supp:eq:self_energy_def} evaluated with the principal logarithm. To examine its behavior near the real axis, write $z=\omega+iy$, with $\ir<\omega<\uv$. In the upper half-plane, we have
\begin{align}
    \Sigma_{\mathrm{I}}(\omega + iy) = \frac{\gamma_0}{2\pi}\Big[ \log(\omega + iy - \ir) - \log(\omega + iy - \uv) \Big].
\end{align}
For the first logarithm, $\omega-\ir>0$, so its argument does not cross the branch cut as $y$ changes from positive to negative values. For the second logarithm, however, $\omega-\uv<0$, so its argument crosses the negative real axis. Applying our logarithm convention on either side gives
\begin{align}
    \begin{dcases}
        \Sigma_{\mathrm{I}}(\omega + i0^+) &= \frac{\gamma_0}{2\pi} \log\left(\frac{\omega - \ir}{\uv - \omega}\right) - i\frac{\gamma_0}{2}, \\
        \Sigma_{\mathrm{I}}(\omega - i0^+) &= \frac{\gamma_0}{2\pi} \log\left(\frac{\omega - \ir}{\uv - \omega}\right) + i\frac{\gamma_0}{2}.
    \end{dcases}
\end{align}
The value below the real axis thus exceeds the value above it by $i\gamma_0$. To remove this discontinuity, we define the analytic continuation for negative imaginary parts as
\begin{align}
    \Sigma_{\mathrm{II}}(z) = \Sigma_{\mathrm{I}}(z)-i\gamma_0
    \qquad \operatorname{Im}z < 0,\quad \ir<\operatorname{Re}z<\uv.
    \label{supp:eq:self_energy_continuation}
\end{align}
On the real axis between the cutoffs, we define $\Sigma_{\mathrm{II}}$ by its limit from below:
\begin{align}
    \Sigma_{\mathrm{II}}(\omega)
    =\Sigma_{\mathrm{I}}(\omega-i0^+)-i\gamma_0
    =\Sigma_{\mathrm{I}}(\omega+i0^+).
\end{align}
The two expressions therefore have the same limit on the real axis. For simplicity, we use $\Sigma(z)$ from now on to denote the resulting analytic function:
\begin{align}
    \Sigma(z) = \begin{dcases}
        \Sigma_{\mathrm{I}}(z), & \operatorname{Im}z>0,\\
        \Sigma_{\mathrm{II}}(z), & \operatorname{Im}z \leq 0.
    \end{dcases}
\end{align}
Figure~\ref{supp:fig:self_energy_continuation} visualizes this continuation inside the frequency band.
\begin{figure}[tb]
    \centering
    \includegraphics[width=176mm]{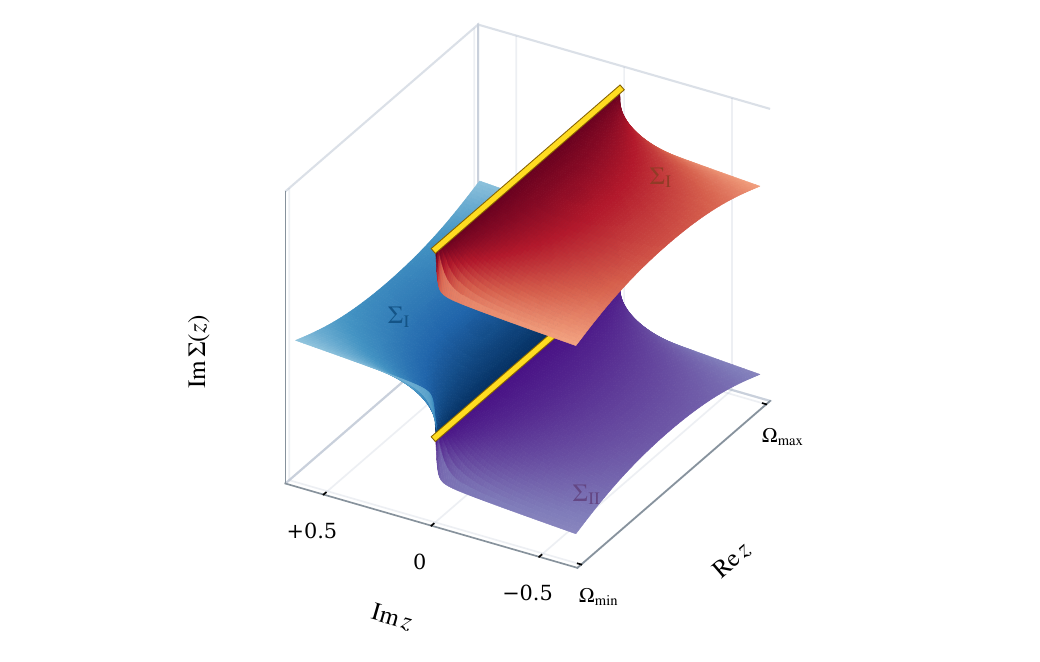}
    \caption{Analytic continuation inside $\Omega_{\mathrm{min}}<\operatorname{Re}z<\Omega_{\mathrm{max}}$, shown through $\operatorname{Im}\Sigma(z)$. The principal branch is blue for $\operatorname{Im}z>0$ and red for $\operatorname{Im}z<0$. The purple surface is $\Sigma_{\mathrm{II}}=\Sigma_{\mathrm I}-i\gamma_0$; it joins the blue surface smoothly across the real axis. Yellow lines mark the two principal boundary values at $\operatorname{Im}z=0$.}
    \label{supp:fig:self_energy_continuation}
\end{figure}
With this definition, the condition $D(z_p)=0$ can be written as the self-consistent equation
\begin{align}
    \Omega_{\mathrm{eff}} - \Omega_0 - i \frac{\gamma_{\mathrm{eff}}}{2} = \Sigma \left(\Omega_{\mathrm{eff}} - i \frac{\gamma_{\mathrm{eff}}}{2}\right),
    \label{supp:eq:self_consistent_pole}
\end{align}
Denoting the left-hand side by $\lambda_p=z_p-\Omega_0$, this becomes
\begin{align}
    \lambda_p = \Sigma(\Omega_0 + \lambda_p), \qquad \lambda_p = \Omega_{\mathrm{eff}} - \Omega_0 - i\frac{\gamma_{\mathrm{eff}}}{2}
\end{align}
Since the self-energy has been properly analytically continued, we can expand the right-hand side around $\Omega_0$, assuming $\ir<\Omega_0<\uv$:
\begin{align}
    \Sigma(\Omega_0+\lambda_p)
    =\sum_{n\geq0}\frac{\Sigma^{(n)}(\Omega_0)}{n!}\lambda_p^n.
\end{align}
Following the analytical continuation prescription, the value at $z = \Omega_0$ is given by
\begin{align}
    \Sigma(\Omega_0)
    &=\Sigma_{\mathrm{I}}(\Omega_0-i0^+)-i\gamma_0 = \frac{\gamma_0}{2\pi}\log\left(\frac{\Omega_0-\ir}{\uv-\Omega_0}\right)-i\frac{\gamma_0}{2}.
\end{align}
This is also the upper limit $\Sigma_{\mathrm{I}}(\Omega_0+i0^+)$. Then, successive differentiation give, for $n\geq1$,
\begin{align}
    \Sigma^{(n)}(z)
    =\frac{\gamma_0}{2\pi}(-1)^{n-1}(n-1)!
    \left[\frac{1}{(z-\ir)^n}-\frac{1}{(z-\uv)^n}\right].
\end{align}
The nearest logarithmic singularities are at the two cutoffs, so this expansion converges provided that
\begin{align}
    |\lambda_p|<\min(\Omega_0-\ir,\uv-\Omega_0).
    \label{supp:eq:pole_series_convergence}
\end{align}
Substituting the Taylor expansion into the pole equation and defining $ \displaystyle \alpha_n=\frac{\Sigma^{(n)}(\Omega_0)}{n!}$ yields
\begin{align}
    \lambda_p=\sum_{n\geq0}\alpha_n(\lambda_p)^n,
    \label{supp:eq:self_consistent_series}
\end{align}
with
\begin{align}
    \begin{dcases}
    \alpha_0&=\frac{\gamma_0}{2\pi}\log\left(\frac{\Omega_0-\ir}{\uv-\Omega_0}\right)
        -i\frac{\gamma_0}{2},\\
    \alpha_n&=\frac{\gamma_0}{2n\pi}
        \left[\frac{1}{(\uv-\Omega_0)^n}
        +\frac{(-1)^{n-1}}{(\Omega_0-\ir)^n}\right],\qquad n\geq1.
    \end{dcases}
\end{align}
This is the self-consistent pole equation stated in the main text. Keeping only the constant and linear terms gives
\begin{align}
    \lambda_p\simeq\frac{\alpha_0}{1-\alpha_1},\qquad
    z_p\simeq\Omega_0+\frac{\alpha_0}{1-\alpha_1}.
\end{align}
Since $\alpha_1$ is real, identifying the real and imaginary parts of $z_p=\Omega_{\mathrm{eff}}-i\gamma_{\mathrm{eff}}/2$ gives
\begin{subequations}
\begin{align}
    \Omega_{\mathrm{eff}}&\simeq\Omega_0-
        \frac{\gamma_0}{2\pi(1-\alpha_1)}
        \log\left(\frac{\uv-\Omega_0}{\Omega_0-\ir}\right),\\
    \gamma_{\mathrm{eff}}&\simeq\frac{\gamma_0}{1-\alpha_1}.
\end{align}
\end{subequations}
These are the effective-parameter relations of the main text, obtained at first order in the self-energy expansion. The decaying solution at this order requires $1-\alpha_1>0$, and the resulting pole must remain within the convergence domain in Eq.~\eqref{supp:eq:pole_series_convergence}.

\section{Relation with the effective amplitude}
\label{supp:sec:effective_amplitude}

For a photon incident with momentum $p=\omega>0$ within the retained band, the reflection amplitude is given, in the notation of the main text, by
\begin{align}
    r^{\mathrm{eff}}_{\omega}
    =\mathcal{T}_{\ir,\uv}(-p,p;\omega)
    =-i\frac{\gamma_0}{2}G_{11}(\omega+i0^+).
    \label{supp:eq:T_pole_decomposition}
\end{align}
Using the pole decomposition~\eqref{supp:eq:pole_decomposition}, with $z_p=\Omega_{\mathrm{eff}}-i\gamma_{\mathrm{eff}}/2$, we obtain
\begin{align}
    r^{\mathrm{eff}}_{\omega}
    =-\frac{i\gamma_0/[2D'(z_p)]}{\omega-\Omega_{\mathrm{eff}}+i\gamma_{\mathrm{eff}}/2}
    -i\frac{\gamma_0}{2}G_{11,\mathrm{reg}}(\omega+i0^+).
    \label{supp:eq:reflection_pole_decomposition}
\end{align}
This decomposition highlights two effects of bandwidth truncation:
\begin{enumerate}[label=(\roman*)]
    \item it modifies the resonant contribution through the \textbf{position} and \textbf{residue} of the pole;
    \item it introduces a regular contribution that may distort the resonance profile.
\end{enumerate}
Matching the pole to the physical parameters is therefore necessary, but not sufficient in general, to recover the full target response: the residue must also provide the correct normalization, while the regular contribution must remain negligible over the relevant frequency range. Within the linear approximation to the self-energy used in the main text, these conditions are simultaneously satisfied. The pole residue acquires the correct normalization, while the regular contribution is neglected around the pole of the projected Green function, yielding
\begin{subequations}
\begin{align}
    \frac{\gamma_0}{2D'(z_p)} &\simeq \frac{\gamma_0}{2(1-\alpha_1)} \simeq \frac{\gamma_{\mathrm{eff}}}{2}, \\
    r^{\mathrm{eff}}_{\omega} &\simeq -\frac{i\gamma_{\mathrm{eff}}/2} {\omega-\Omega_{\mathrm{eff}}+i\gamma_{\mathrm{eff}}/2}.
\end{align}
\end{subequations}
The bandwidth-reduction strategy therefore targets an intermediate regime in which the cutoffs are close enough to induce appreciable shifts of the effective parameters, yet sufficiently far from resonance for the self-energy to remain approximately linear over the relevant frequency range. As the cutoffs approach the resonance, higher-order contributions become significant and the above conditions cease to hold, such that the truncated simulation can no longer be fully corrected.

\nocite{pletyukhovScatteringMasslessParticles2012,
  chenPropagatorsJaynesCummingsModel2002,
  seeDiagrammaticApproachMultiphoton2017,
  shiMultiphotonScatteringTheory2015,
  sheremetWaveguideQuantumElectrodynamics2022,
  stefanoFeynmandiagramsApproachQuantum2017,
  rouletTwoPhotonsAtomic2016,
  peskinIntroductionQuantumField1995,
  collinsRenormalizationIntroductionRenormalization1984,
  schneiderGreensfunctionFormalismWaveguide2016,
  scullyQuantumOptics1997,
  domokosQuantumDescriptionLight2002}
\clearpage
\endgroup

\EOD
\end{document}